\documentclass[12pt,document,nofootinbib,superscriptaddress, onecolumn,preprintnumbers,balancelastpage]{revtex4}
\pdfoutput=1
\usepackage{latexsym}
\usepackage{amssymb}
\usepackage{epsfig,amsmath,graphics}
\usepackage{listings}
\usepackage{color}
\usepackage{dcolumn}
\usepackage{slashed}
\usepackage{comment,latexsym} 
\usepackage{bm}
\usepackage{verbatim}
\usepackage{longtable,tabularx}
\usepackage{dcolumn}
\definecolor{Mahogany}{rgb}{0.62,0.24,0.15}
\definecolor{colorLink}{rgb}{0.7,0,0}
\definecolor{colorCite}{rgb}{0,.7,0}
\definecolor{colorURL}{rgb}{0,0,0.7}
\usepackage[colorlinks=true,linktocpage=true,linkcolor=black,citecolor=colorCite,urlcolor=colorURL]{hyperref}
\usepackage{setspace}
\usepackage{xspace}
\usepackage{multirow}
\usepackage{titlesec}

\newcommand{\OO}{\mathcal{O}}

\def\be{\begin{equation}}
\def\ee{\end{equation}}
\newcommand{\beq}{\begin{equation}}
\newcommand{\eeq}{\end{equation}}
\def\bea{\begin{eqnarray}}
\def\eea{\end{eqnarray}}
\newcommand{\eref}[1]{Eq.~(\ref{#1})}

\newcommand{\GeV}{{\text{ GeV}}}

\newcommand{\jargon}[1]{{\bf \color{Mahogany}{#1}}}

\newcommand{\vx}{\vec{x}}
\newcommand{\ith}{i^{\text{th}}}
\newcommand{\pT}{{p_T}}
\newcommand{\trho}{\hat{\rho}}
\newcommand{\srho}{\trho^{~\!\!\star}{}}
\newcommand{\kk}{\text{k}}
\newcommand{\nj}{{N_j}}
\newcommand{\nbs}{{N_\text{\tiny{BS}}}}
\newcommand{\vk}{\vec{\kk}}
\newcommand{\vz}{\vec{z}}
\newcommand{\scinot}[2]{#1\times 10^{#2}}
\newcommand{\dd}{\mathrm{d}}
\newcommand{\sep}{\big |}

\expandafter\def\expandafter\normalsize\expandafter{%
    \normalsize
    \setlength\abovedisplayskip{8pt}
    \setlength\belowdisplayskip{8pt}
    \setlength\abovedisplayshortskip{8pt}
    \setlength\belowdisplayshortskip{8pt}
}

\titleformat{\section}{\center\normalfont\fontsize{14}{15}\bfseries}{\thesection.}{1em}{}
\titleformat{\subsubsection}{\center\normalfont\fontsize{12}{15}}{\thesubsubsection.}{1em}{}

\begin{document}

\begin{flushright}
\text{\normalsize SLAC-PUB-15891}
\end{flushright}
\vskip 45 pt

\title{Jet Substructure Templates: \\Data-driven QCD Backgrounds for Fat Jet Searches} 

\author{$\quad\quad\quad\quad\quad\quad\quad\quad$ Timothy Cohen}
\affiliation{
Theory Group, SLAC National Accelerator Laboratory\\
\vskip -6 pt
Menlo Park, CA, 94025}

\author{Martin Jankowiak}
\affiliation{
Institut f\"ur Theoretische Physik, Universit\"at Heidelberg\\
\vskip -6 pt
69120 Heidelberg, Germany}

\author{$\quad\quad\quad\quad\quad\quad\quad\quad$ Mariangela Lisanti}
\affiliation{
Physics Department, Princeton University\\
\vskip -6 pt
Princeton, NJ 08544}

\author{Hou Keong Lou}
\affiliation{
Physics Department, Princeton University\\
\vskip -6 pt
Princeton, NJ 08544}

\author{Jay G. Wacker}
\affiliation{
Theory Group, SLAC National Accelerator Laboratory\\
\vskip -6 pt
Menlo Park, CA, 94025}

\begin{abstract}
\vskip 15 pt
\begin{center}
{\bf Abstract}
\end{center}
\vskip -30 pt
$\quad$
\begin{spacing}{1.2}

QCD is often the dominant background to new physics searches for which jet substructure provides a useful handle.  Due to the challenges associated with modeling this background, data-driven approaches are necessary.  This paper presents a novel method for determining QCD predictions using templates -- probability distribution functions for jet substructure properties as a function of kinematic inputs.  Templates can be extracted from a control region and then used to compute background distributions in the signal region. Using Monte Carlo, we illustrate the procedure with two case studies and show that the template approach effectively models the relevant QCD background.  This work strongly motivates the application of these techniques to LHC data.
 
\end{spacing}
\end{abstract}

\maketitle
\newpage
\begin{spacing}{1.3}
\large
\tableofcontents
\normalsize
\pagebreak
\section{Introduction}
Analyzing the substructure of jets has proven useful for a multitude of new physics scenarios.  A variety of substructure observables and techniques have been proposed \cite{Abdesselam:2010pt,Altheimer:2012mn,Altheimer:2013yza}, with applications ranging from boosted-object tagging \cite{Butterworth:2008iy, Kaplan:2008ie, Plehn:2010st, Cui:2010km} to high-multiplicity searches that take advantage of accidental substructure \cite{Hook:2012fd, Cohen:2012yc, Hedri:2013pvl}.  Simultaneously, tremendous progress has been made in modeling QCD at the analytical level and with Monte Carlo (MC) event generators~\cite{Ellis:2009zw,Berger:2009zg,Bredenstein:2009aj,Bevilacqua:2009zn,Binoth:2009rv,Ita:2011wn,Alwall:2007st,Mangano:2002ea,Gleisberg:2008ta,Alwall:2007fs,Catani:2001cc,Hirschi:2011pa, Hoche:2010pf, Alioli:2010xd, Nason:2012pr, Nason:2004rx, Frixione:2002ik,Dasgupta:2012hg,Chien:2012ur,Jouttenus:2013hs,Rubin:2010fc,Feige:2012vc,Larkoski:2013eya,Dasgupta:2013via,Larkoski:2013paa}.  These developments have been crucial in supporting efforts to perform Standard Model measurements, as well as searches for new physics, at the LHC.  However, given the complexity of the LHC environment, data-driven background estimates for QCD remain a central component of any jet substructure study; new methods for determining these backgrounds are thus of great interest.

There are three complementary reasons why jet substructure is useful for new physics searches at the LHC.  The first is that large-radius ``fat" jets \cite{Seymour:1993mx} naturally accommodate grooming methods like trimming \cite{Krohn:2009th}, pruning \cite{Ellis:2009me}, and mass-drop filtering \cite{Butterworth:2008iy}.  This is a significant advantage in the hadronic environment of the LHC, where the presence of the underlying event and pile-up makes mass reconstruction difficult. Second, teasing out the signatures of highly boosted states that undergo hadronic decays is difficult to impossible without jet substructure techniques.  Maximizing sensitivity to very boosted final states will become increasingly important as rising trigger thresholds make low-$\pT$ regions of phase space inaccessible.   Finally, the fat jet paradigm provides an efficient way to separate signal from background in multijet events.  For example, the total jet mass of an event (the sum of the individual fat jet masses) is a particularly useful discriminator~\cite{Hook:2012fd} due to the simple fact that QCD does not tend to yield large-mass jets.  

Given the many successes of the fat jet approach and the practical importance of improving methods for obtaining data-driven estimates for Standard Model rates, it is natural to focus on the specific case of the multijet background to fat jet searches.  In \cite{Hedri:2013pvl, Bai:2013xla}, a novel approach to computing multijet backgrounds was proposed that relies on the assumption that the properties of individual fat jets are independent of one another and universal -- this is motivated by the approximate factorization of QCD jets~\cite{Collins:1989gx}.   Under this assumption, jet properties can be measured in signal-poor samples (\emph{e.g.}, inclusive dijets) and then extrapolated to the signal (\emph{e.g.}, multijet) region.  This paper provides a concrete procedure for making such an extrapolation.
 
Typically, data-driven techniques require identifying a set of approximately uncorrelated variables to define control regions.  Any non-negligible correlations require correction factors that are estimated with MC.  Consequently, if one wants to perform cuts on non-trivially correlated variables, such as the mass and other substructure properties of a jet, the background estimate is no longer data-driven.  In contrast, the template methods introduced in this paper naturally incorporate these correlations.  As a result, searches can be performed with more handles than previously possible and with less dependence on MC. 

Any realistic data-driven proposal must assess the statistical and systematic uncertainties associated with the background estimate.  By taking advantage of kernel smoothing techniques, we develop a framework in which the statistical uncertainties associated with the template method can be reliably addressed.  We demonstrate the feasibility of our approach by applying it to two example mock analyses.  

The rest of the paper is organized as follows. Sec.~\ref{sec:Approach} provides a general overview of the method.  Sec.~\ref{sec:Techniques} continues the discussion with  a more detailed and technical description. Sec.~\ref{sec:Applications} presents two MC case studies to demonstrate how the method might be deployed in a realistic analysis.  Finally, Sec.~\ref{sec:Conclusions} presents our conclusions.   An Appendix is included that gathers together many of the technical aspects of this study.  A review of kernel smoothing is included in App.~\ref{app:KernelSmoothing}.  The technical details of our MC generation and processing are discussed in App.~\ref{sec:MCFramework}.  Code for implementing the approach presented in this paper 
is available at \texttt{http://sourceforge.net/projects/kernelsmoother/}.
 
To keep track of the terminology associated with our proposal, new terms are emphasized with \jargon{bold red} typeface where they are first introduced.

\section{Substructure Templates}
\label{sec:Approach}

The approach advocated here is different from typical data-driven methods.  In broad strokes, the usual strategy is to 
\begin{enumerate}
\vspace{-5pt}
\item  define a \jargon{control region} that is expected to be signal-poor;
\vspace{-10pt}
\item  use this region to validate/normalize MC calculations; 
\vspace{-10pt}
\item use the validated MC to predict the backgrounds in the \jargon{signal region} for which a given search has been optimized.
\end{enumerate}
The foundation of our approach, in contrast, involves constructing a data-driven \jargon{template} that models the substructure of fat jets.  This template is then used to \jargon{dress} a sample of multi-fat jet events.{\interfootnotelinepenalty=10000\footnote{Note that other examples of ``dressing" have been implemented in the past.  For example, in the context of charged Higgs boson searches, ATLAS makes a data-driven background estimate by taking a control sample with muons in the final state and replacing the muons with simulated taus, a technique that ATLAS refers to as ``embedding"~\cite{ATLAS-CONF-2011-051}.  This is essentially the inversion of our recipe.  CMS has also used a version of dressing in searches for boosted $t\,\overline{t}$ resonances.  In particular, they rely on data in a control region to determine the properties of their top tagger, which is then propagated to the signal region~\cite{Chatrchyan:2012ku, Chatrchyan:2013lca}.}}  The selection criteria are then applied to the dressed sample to yield a background estimate in the signal region.  

\subsection{Separating Kinematics and Substructure}
The templates $\hat{\rho}$ are transfer functions derived from data that encode the statistical distribution of output variables as a function of the input variable(s).  In the limit of infinite statistics, $\hat{\rho}$ converges to the true distribution $\rho$.  The outputs are jet substructure observables, such as the jet mass, while the inputs are kinematic variables. Assuming that the input variable is the jet's transverse momentum  $\pT$, then
\begin{eqnarray}
\hat P\big(m\,\sep\, \pT\big) = \trho\big(m \, \sep\, \pT\big)\,\mathrm{d}m
\end{eqnarray}
is the data-driven estimate for the probability $\hat P$ that a jet with the given $p_T$ has a mass between $m$ and $m+\mathrm{d}m$.  Here, ``$\sep$" distinguishes input from output variables on the right and left sides, respectively, and hatted variables refer to estimated quantities.

Templates can be used to compute data-driven backgrounds through the following procedure:
\begin{enumerate}
\vspace{-5pt}
\item determine a control region that is expected to be signal-poor;
\vspace{-10pt}
\item map out templates using data in the control region;
\vspace{-10pt}
\item generate a MC sample in the signal region, disregarding everything except for the distribution of the input variable(s);
\vspace{-10pt}
\item convolve the templates derived from the control sample with the MC sample;
\vspace{-10pt}
\item apply cuts to this data/MC hybrid dataset to obtain a data-driven background estimate.
\end{enumerate}

This article develops a concrete formalism for determining QCD backgrounds in this way, with particular care being paid to the statistical properties of the results.  In more detail, the procedure is formulated as follows.  First, a suitable control region is defined, which is referred to as the \jargon{training sample}. The fat jets in this sample are used to construct a histogram for the substructure observable(s) of interest as a function of kinematic observable(s).  In order to obtain a continuous template from a limited number of fat jets, the histogram must be smoothed.  This is done using kernel smoothing techniques, which are described in detail in Sec.~\ref{sec:Techniques} and App.~\ref{app:KernelSmoothing}.  The statistical error associated with the kernel smoothing procedure is folded into the final error estimate for the number of background events surviving a given set of cuts.
\begin{figure}[t] 
   \centering
   \includegraphics[width=5 in]{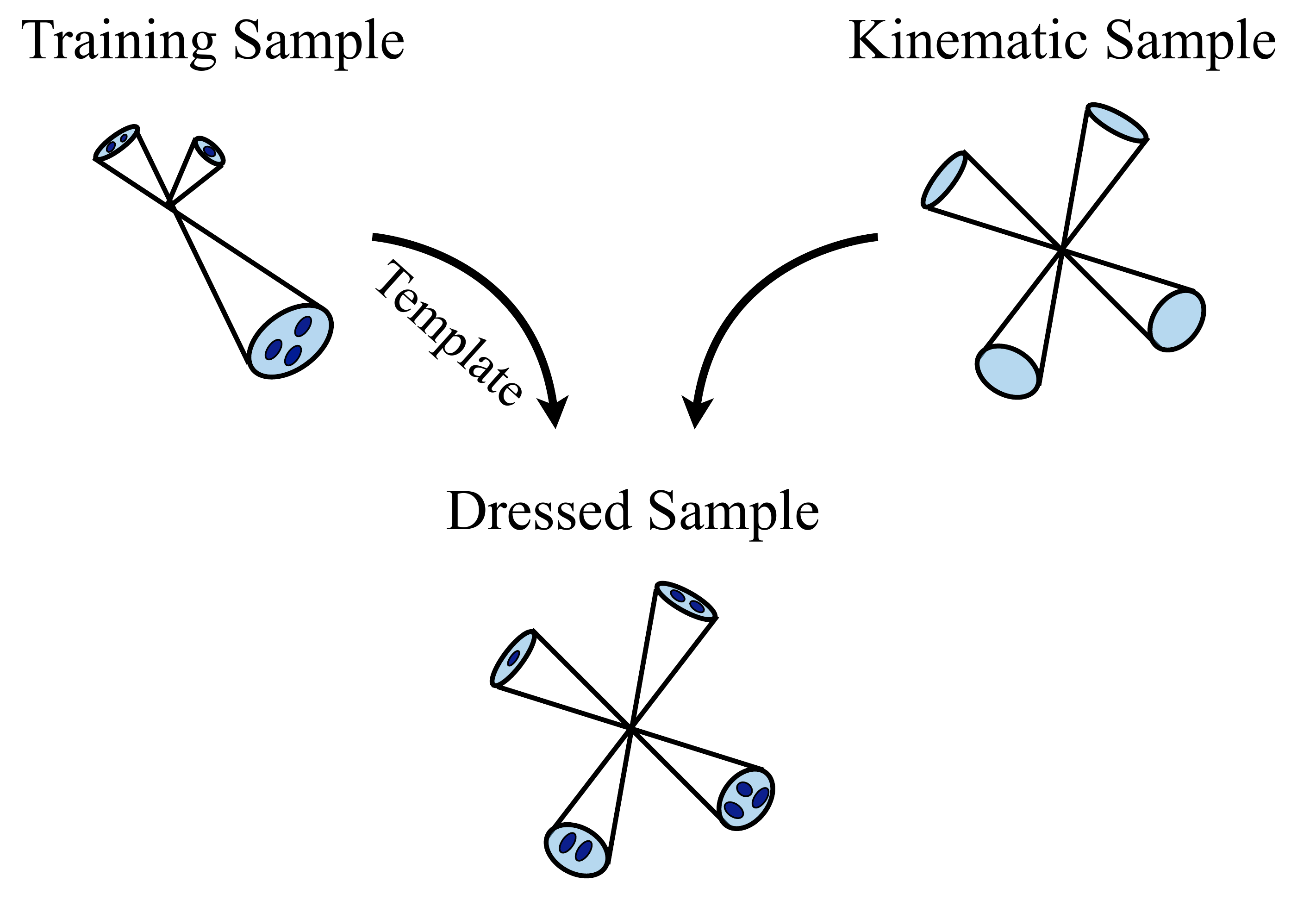} 
   \caption{A pictorial representation of the procedure.}
   \label{fig:example}
\end{figure}

The next step requires the use of MC integration and is the analog of the extrapolation performed in other data-driven approaches.  To begin, a \jargon{kinematic sample} is generated with event generator MC.  The kinematic sample consists of events with $\nj$ fat jets.  This sample is then used to extract the kinematic distribution of the fat jet background -- \emph{e.g.},
\be
\frac{\text{d}^\nj\sigma({\pT}_i)}{\text{d}{\pT}_1 ... \text{d}{\pT}_{\nj}}  \,\, .
\ee
It is worth emphasizing that any jet substructure information in the kinematic sample goes unused.  Instead, each fat jet in the kinematic sample is dressed with substructure information using the template determined from the previous step; we refer to this as Monte Carlo integration, which results in the \jargon{dressed sample}.  Lastly, cuts are performed on the dressed data set to obtain a background estimate and an associated smoothing variance.  Note that this approach incorporates the correlations between the kinematics of the various fat jets into the final result.  As shown in Sec.~\ref{sec:Applications} using two explicit mock analyses, this allows non-trivial correlations to propagate to the dressed sample and reproduces the predictions from MC, within statistical uncertainties.

To summarize, the proposed data-driven strategy is (this parallels the enumeration above):
\begin{enumerate}
\vspace{-5pt}
\item determine a control region to obtain a training sample;
\vspace{-10pt}
\item train a template $\trho$;
\vspace{-10pt}
\item generate a kinematic sample using MC;
\vspace{-10pt}
\item perform the integration thereby convolving $\trho$ with the kinematic sample; 
\vspace{-10pt}
\item apply cuts to the resulting set of dressed events to obtain a data-driven background estimate.
\end{enumerate}
Fig.~\ref{fig:example} provides a pictorial representation of the procedure.  The computation of smoothing variance is more involved and we postpone a detailed discussion to Sec.~\ref{sec:Techniques}. 
\subsection{Using Jet Factorization}
\label{sec:Factorizability}
Let us now take a moment to discuss the assumption of jet template independence in more detail.  This assumption leads to a dramatic reduction in the dimension of the multijet phase space and allows for data-driven background estimates to be made with minimal statistical error. 

The jet properties we study here can be separated into two categories.  Those in the first category, denoted by the $d$-dimensional vector $\vx$, are the outputs of templates.    Examples include jet mass, $N$-subjettiness, angularity, and other jet substructure observables.  The physics of these observables is largely driven by the parton shower and, as such, is set by the kinematics of the hard parton. Physically, the corresponding templates can be mapped out by taking a parton-level event and showering it many times.  The $d$ observables in $\vx$ need not be independent of each other, and in most cases they are not.  For instance, if a jet's mass is anomalously large, the same jet will have non-trivial $N$-subjettiness moments. 

The second category consists of kinematic variables, denoted by $\vk$.  In this paper, we only consider the jet $\pT$, although in principle the set $\vk$ could be augmented with other variables, such as the pseudo-rapidity of the jet or pileup conditions at the time of the event.  These kinematic properties cannot be modeled as template outputs, since the kinematics of jets in an event are strongly correlated with one another.    

An $\nj$-jet differential cross~section is given as a function of the fat jet kinematics $\vk_i$ and substructure observables $\vx_i$ for $i=1$ to $\nj$.  It can be written as\footnote{Note that $\rho(\vx_i \sep \vk_i)$ is a legitimate (conditional) probability distribution, \emph{i.e.}~for each fixed value of the $\vk_i$ the distribution $\rho(\vx_i \sep \vk_i)$ yields unity 
when integrated over the $\vx_i$. }
\be
\label{eqn:factorization}
\frac{\dd^{2\nj} \sigma(\vx_i,\vk_i, )}{\dd \vx_1 ... \dd \vx_\nj \dd \vk_1 ... \dd \vk_\nj \; } = \frac{\dd^\nj \sigma(\vk_i)}{\dd\vk_1 ... \dd \vk_\nj} \; \rho(\vx_1, ..., \vx_\nj \sep \vk_1, ..., \vk_\nj) \,\,.
\ee
For a given $\vk_i$, we assume that the substructure observable $\vx_i$ of the $i^{\rm{th}}$ fat jet is independent of all the other $\vx_j$ and $\vk_j$, with $j \ne i$.  Then the right-most term in \eref{eqn:factorization} splits into a product of $\nj$ terms:\be
\label{eqn:factorization2}
\frac{\dd^\nj \sigma(\vk_i)}{\dd\vk_1 ... \dd\vk_\nj} \; \rho(\vx_1, ..., \vx_\nj \sep \vk_1, ..., \vk_\nj) = \frac{\dd^\nj \sigma(\vk_i)}{\dd\vk_1 ... \dd\vk_\nj} \prod_{i=1}^{\nj} \rho_i(\vx_i \sep \vk_i) \, .
\ee
This factorized form of the differential cross section allows for background estimates to be computed by a convolution of separate kinematic and substructure parts and forms the basis of our data-driven proposal.  Note that the independence assumption underlying \eref{eqn:factorization2} dramatically reduces the dimensionality of the configuration space that needs to be explored. Consider, for example, the case where $d=1$ and $\kk=\pT$. Then, instead of using MC to estimate the QCD prediction for a $2\nj$-dimensional configuration space, only a $\nj$-dimensional configuration space needs to be explored. If multiple jet properties are being studied, $d>1$, then the reduction in dimensionality is even more dramatic, 
with a $(d+1)\times \nj$ dimensional configuration space reduced to $\nj$ dimensions.  The additional cost involved is creating the $(d+1)$-dimensional templates $\rho(\vx \sep \pT)$. Of course, as $d$ increases (keeping the size of the training sample fixed), the density estimation that underlies the construction of substructure templates becomes increasingly difficult and at some point the associated uncertainties will outweigh any gains from the reduced dimensionality.  This is the price to pay for not including any \emph{a priori} model for correlations between the various substructure observables; however, these minimal assumptions allow us to remain data-driven.

Soft QCD effects can violate the factorization in \eref{eqn:factorization2}.  These include underlying event, pile-up, and $\mathcal{O}(N_c^{-2})$ corrections coming from perturbative QCD due to large-angle soft radiation.  Jet grooming can minimize these effects.  Throughout this paper, the fat jets are trimmed~\cite{Krohn:2009th}, as described in detail in App.~\ref{sec:MCFramework}.  In general, it is important to consider the conditions under which \eref{eqn:factorization2} is most likely to hold -- for example, as a function of the fat jet radius.

There are many additional subleading effects that spoil the factorization in \eref{eqn:factorization2}. 
The most important of these is the quark-gluon composition of jets in an event.  If the composition of the jets changes dramatically between the training sample and the signal region, then anomalous features may be introduced.  In practice, quark- and gluon-initiated jets often have broadly similar features in the signal region and the composition does not change dramatically in many samples of interest.  However, the templates taken from the two leading jets do not describe the third and fourth jets well.  The introduction of quark and gluon templates may resolve this issue, as will be explored in future work~\cite{FutureWork}.

Note that the three-momenta of the fat jets in the dressed data are automatically conserved because they are taken from MC.  However, energy is not conserved since the fat jet masses are determined stochastically -- the total energy of each dressed event is different from that of the kinematic event.  In practice, we do not find that this is an issue for the cases studied here.  The violation of energy conservation changes the center-of-mass energy of the event, but not appreciably because the ratio $m_j / \pT < 1$.  These effects lead to correlations between the masses of the jets; how to account for this in the template approach remains an interesting open question.

\section{Implementing Templates}
\label{sec:Techniques}

This section provides a detailed description of the procedure for generating substructure templates and calculating cut efficiencies.  It is accompanied by App.~\ref{app:KernelSmoothing}, which reviews the basics of the statistical procedures used here.  If the reader is new to kernel smoothing methods, we recommend that he or she take a moment to review App.~\ref{app:KernelSmoothing} before proceeding.

\subsection{Constructing Templates}

The jets in the training sample are used to build a histogram.  Technically, this is all that is needed to build the templates.  However, due to the finite size of the training sample, the histograms might not be adequately populated and data-smoothing techniques are necessary.

\begin{figure}[t!]
\centering
\includegraphics[width=1.\textwidth]{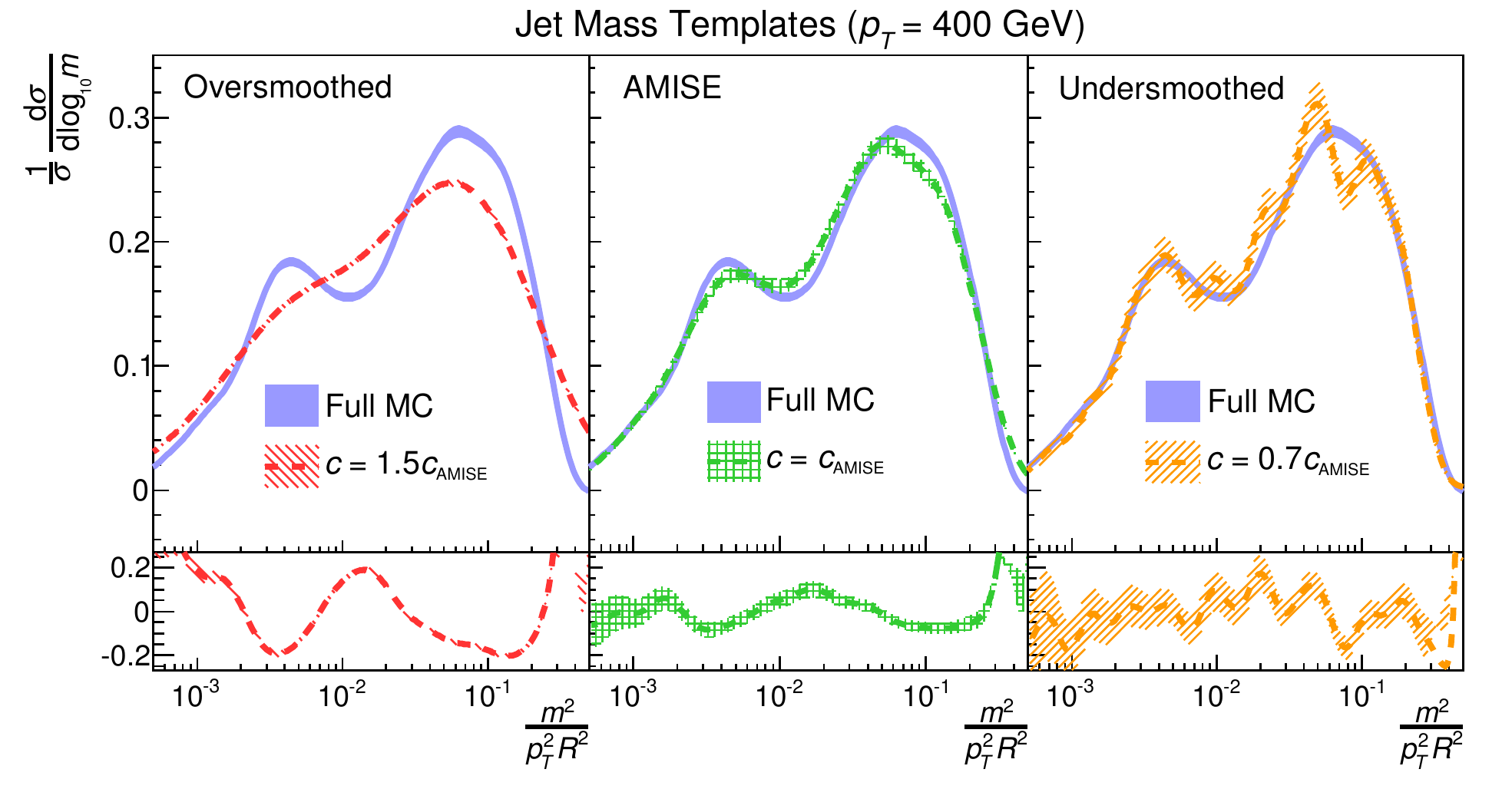}
\caption{The $p_T = 400 \GeV$ jet mass template is shown for three distinct bandwidths.  The error band indicates the variance, while the bias can be seen in the mismatch between the solid curve and the dashed curves (variance and bias are defined later in this section).  The lower panels give the fractional deviation between the solid and dashed curves.  }
\label{fig:ExampleTemplate}
\end{figure}

The statistical technique of kernel smoothing is one way to approach the problem.  In this procedure, every data point in the training sample is smoothed out to the adjoining region in the multi-dimensional space.  The aggregate of these smooth contributions gives the final probability distribution function.  Take, as an example, the jet mass distribution (solid purple) shown in Fig.~\ref{fig:ExampleTemplate}.  This distribution is derived by looking at the two leading jets with $p_T > 200 \text{ GeV}$ in the full MC sample.  Now imagine that we want to reproduce this distribution using a statistically limited sample -- say 1\% of the total number of events in the full MC sample.  Applying kernel smoothing to the statistically limited sample reproduces the full distribution, to within errors.

Take an event with $N_j$ jets, each with particular kinematic and substructure properties described by the vectors $\vk$ and $\vx$, respectively.  For example, if we are interested in each jet's $p_T$, mass, and $N$-subjettiness ratio $\tau_{21}$, then the $i^{\text{th}}$ jet (ordered by $p_T$) is described by $\vk_{j_i} = \left(p_{T}\right)_{j_i}$ and $\vx_{j_i}=\left ( m, \tau_{21}\right)_{j_i}$.  We unify $\vk$ and $\vx$ into the single $D$-dimensional vector $\vz$, where $D$ is the number of kinematic and substructure variables of interest.  Then, the $i^\text{th}$ jet in event $j$ is described by  
\begin{eqnarray}
J_{ij}\big[\vz_{\,}\big] \equiv J_{ij}\Big[\vx, \vk_{\,}\Big] .
\end{eqnarray}

To build a \jargon{training sample} for the $\ith$ jet, we combine the $\ith$ jet from \emph{all} events in the sample.  Said another way, the training sample for the $\ith$ jet, denoted as $T_{i}$, is the set
\begin{eqnarray}
T_{i} = \Big\{ J_{i1}\big[\vz_1\big] \,, J_{i2}\big[\vz_2\big] \,, \ldots \,, J_{i\,N_T}\big[\vz_{N_T}\big] \Big\} \, , 
\end{eqnarray}
where $N_T$ is the number of events with an $\ith$ jet.  
The normalized histogram corresponding to $T_i$ encodes a discrete probability distribution function with $D$ dimensions.  This probability distribution function may or may not be smooth, depending on the total number of events $N_T$.  This is where kernel smoothing comes into play -- it allows us to define a smooth probability distribution function in a statistically robust manner.  In particular, we can define the smoothed template for the $\ith$ jet, $\trho_i\big(\vz_{\,}\big)$, using 
\be
\label{eq:DefinePDF}
\trho_i\big(\vz_{\,}\big)=\frac{1}{N_T}\sum_{J\in T_i} K_h\big(\vz-\vz(J)\big)
\qquad \text{and}\qquad
K_h\big(\vz_{\,} \big) = \frac{1}{\det h}K\big(h^{-1} \vz_{\,} \big),
\ee
where $\vz(J)$ is the value of $\vz$ for jet $J$, $K$ is the \jargon{kernel}, and $h$ is a matrix describing the kernel width.  The choice of kernel is not too critical and we simply adopt the canonical choice of a Gaussian:
\be
K(\vec z)=\frac{1}{(4\pi)^{\frac{D}{2}}}\exp
\left(-\frac{\vert\vec z\,\vert^2}{2}\right) 
\quad \text{and} \quad
K_h(\vec z)=\frac{1}{(4\pi)^{\frac{D}{2}}\det h}\exp
\left[-\Big(2\, h^T h\Big)^{-1}_{ij}z^i z^j\right] \,\, .
\ee

The template generated by the kernel smoothing procedure, \eref{eq:DefinePDF}, is simply a sum of Gaussians around each data point in the training sample.  The degree to which this procedure appropriately smooths out the distribution depends on the matrix of widths, $h_{ij}$.  Thus, the goal is to choose the smoothing parameters such that the estimated distribution function, $\trho\big(\vz_{\,}\big)$, is as close as possible to the true distribution function, $\rho\big(\vz_{\,}\big)$.  Selecting the width matrix $h_{ij}$ of the kernel is one of the most important aspects of this procedure because an over/under-smoothed distribution may not accurately represent the true distribution function.  A common practice is to use the ``Asymptotic Mean Integrated Squared Error'' (AMISE) metric described in App.~\ref{app:KernelSmoothing}.  This gives rise to ``Silverman's Rule-of-Thumb" \cite{Silverman86}, where $h_{ij}$ is 
\be
\label{eq:SilvermanRule}
h^{\text{\tiny{AMISE}}}_{ij}\simeq c \,\hat\sigma_{ij} N_T^{-\frac{1}{D+4}} \,\, .
\ee
Here, $c$ is an $\OO(1)$ constant 
and $\hat\sigma_{ij}$ is an estimator for the square root of the covariance matrix of $\rho$.  For distributions that are not vastly different from the normal distribution, this provides a useful starting point for determining the kernel width. 

Figure~\ref{fig:ExampleTemplate} illustrates the effects of choosing different values of $c$.  All three panels show the result of applying kernel smoothing to the full MC sample; this serves as an approximation to the ``true" distribution.\footnote{The non-trivial shape of these distributions results from jet trimming, see \cite{Dasgupta:2013ihk}.}  Overlaid are templates that have been generated using only 1\% of the full sample in order to demonstrate how kernel smoothing approximates the true distribution for different values of the kernel width.{\interfootnotelinepenalty=10000\footnote{It is important to mention one subtlety in this figure.  The solid curve is the corrected jet mass template $\srho$ (see \eref{eq:CorrectedTemplate}).  The dashed curves are the uncorrected templates $\trho$ (see \eref{eq:DefinePDF}) and demonstrate the bias inherent in the smoothing procedure.}}  The width is varied from the optimal value $c = c_\text{\tiny{AMISE}}$ [middle, green] to an over-smoothed value $c = 1.5 \, c_\text{\tiny{AMISE}}$ [left, red], and an under-smoothed value $c = 0.7 \,c_\text{\tiny{AMISE}}$ [right, orange]. 

The smoothing procedure introduces two competing sources of error in the template, \jargon{bias} and \jargon{variance}, whose
magnitudes are set by the amount of smoothing.  These are defined as follows: 
\begin{eqnarray}
\text{bias} &\equiv&  b\big(\vz_{\,}\big) =  \rho\big(\vz_{\,}\big) - \trho\big(\vz_{\,}\big) \,; \\
\vspace{-10pt}
\text{variance} &\equiv& v^2\big(\vz_{\,}\big) = \Big\langle \trho\big(\vz_{\,}\big)^2 \Big\rangle - \Big\langle \trho\big(\vz_{\,}\big)\Big\rangle^2 \, ,
\label{eq:DefOfVar}
\end{eqnarray}
where $\langle ... \rangle$ is the expectation value with respect to the true distribution $\rho$.
The variance is interpreted as the scatter in $\trho\big(\vz_{\,}\big)$ when estimated from different samples drawn from $\rho\big(\vz_{\,}\big)$, while the bias is a measure of the systematic displacement of the estimated distribution from the true distribution as a consequence of smoothing.

As discussed in more detail in App.~\ref{app:KernelSmoothing}, the presence of bias is problematic and must be corrected for. Throughout, we use a \jargon{corrected template} $\srho$, which has substantially reduced bias.  To compute $\srho(\vz)$, all that
is needed is an estimate for $b(\vz)$. We adopt an estimator $\hat b\big(\vz_{\,}\big)$ that is formed by comparing the twice-smoothed template 
\be
\hat{\hat{\rho}}\big(\vz_{\,}\big) = \int \dd^D z^{\prime} \,\hat\rho\big(\vz^{\,\,\prime} \big) K_h\big(\vz - \vz^{\,\prime}\big) 
\ee
with the once-smoothed template $\trho\big(\vz_{\,}\big)$: 
\be
\hat b\big(\vz_{\,}\big) = \hat{\hat{\rho}}\big(\vz_{\,}\big) - \trho\big(\vz_{\,}\big).
\label{eq:Bias}
\ee
This estimator is valid to leading order in the kernel width \cite{scott}, as shown in \eref{eq:TemplateBiasImp}.  The corrected template is then defined as
\be
\label{eq:CorrectedTemplate}
\srho\big(\vz_{\,}\big) = \trho\big(\vz_{\,}\big) - \hat b\big(\vz_{\,}\big).
\ee
It is this corrected template that will be used to make background estimates in the signal region.  This procedure is justified in App.~\ref{app:KernelSmoothing}, where the corrected template is shown to give consistent results for a test probability distribution.

\subsection{Dressing an Event}

The previous section showed how to construct a template from a training sample. With this template in hand, one can calculate the efficiency for a kinematic event in the signal region to pass a given set of cuts.  We will now describe this calculation in detail.  
Note that throughout the rest of this paper, unless stated otherwise, our substructure templates are universal.  That is, we do not make any distinctions between, say, the $1^{\rm{st}}$ and $2^{\rm{nd}}$ jets (as ordered by $\pT$) in the training sample. Rather, a single template is constructed making use of all the jets in the training sample.

The corrected template $\srho(\vz)$ constructed in the previous section does not distinguish between the kinematic variables
$\vk$ and substructure variables $\vx$.  However, we wish to view the substructure variables as functionally dependent upon the kinematic input variables.  That is, we want to convert the joint probability distribution $\srho(\vx, \vk)$ into the conditional probability distribution $\srho (\vx\sep \vk)$ using
\be
\srho\big( \vx\sep \vk_{\,}\big) = \frac{\srho\big( \vx, \vk_{\,}\big)}{\srho\big(\vk_{\,}\big)} = \frac{\srho\big( \vx, \vk_{\,}\big)}{\int \!\dd^{d}x' \; \srho\big( \vx^{\,\prime}, \vk_{\,}\big)}
\quad \text{such that} \quad
\int \!\!\dd^{d}x \; \srho( \vx\sep \vk)  = 1 
\ee
for each $\vk$.  The template $\srho( \vx\sep \vk_{\,}) $ is normalized so that it acts as a probability distribution function for the substructure information of a jet with given kinematics $\vk$.

Given a multi-fat jet event in the signal region, obtained from MC, we can apply cuts to the first and second jets, with specified kinematics $\vk_{1}$ and $\vk_{2}$, respectively.  The efficiency, $\hat\epsilon^\star$, that this event passes the cuts is
\begin{eqnarray}
\label{eq:effContinuous}
\hat\epsilon^\star[C] = \int_{C} \, \, \!\!\! \dd^dx_1\, \dd^dx_2\;\;\srho\big(\vx_1\sep \vk_{1}\big) \;\srho\big(\vx_2 \sep \vk_{2}\big) \, ,
\end{eqnarray}
where $C=C\big[\vx_1,\vx_2\big]$ is the region of the $(\vx_1, \vx_2)$ configuration space selected by the cuts.  For instance, if $\vx$ is the mass of a jet, then
\begin{eqnarray}
\label{Eq: CutDef}
C\big[\vx_1,\vx_2\big]=  \{ m_1 + m_2 > M_J \}
\end{eqnarray}
is the cut requiring that the sum of the two individual jet masses is larger than $M_J$.  While the integrand of \eref{eq:effContinuous} is a product of templates, the region of integration may be a complicated function of multiple variables.   

\subsection{Calculating Errors} 
\label{sec:PredictionsWithErrorBars}

This procedure is trivially generalized to a \emph{set} of kinematic events, thus yielding a background prediction with error bars.  
The most straightforward use of the template is to start with a sample of jets, of size $N_e$, where only the kinematics are specified.  A dressed MC sample may be created by dressing each event with the template multiple times.  Said another way, the integral in \eref{eq:effContinuous} is performed by Monte Carlo integration; for each kinematic event, the template is used to produce $n_I$ dressed events. It is sufficient to take $n_I\sim \OO(10^3)$.  Schematically, to dress events with two kinematic objects, one converts each MC event into $n_I$ dressed events 
\be
\label{eq:dressedPT}
\Big(J\big[\vk_1\big],\, J\big[\vk_2\big]\Big)_i \quad \longrightarrow \quad \Big\{\Big(\big(J \big[\vk_1, \vx_{1\alpha}\big], J\big[\vk_2, \vx_{2\alpha}\big]\big)_i, w_\alpha^\star, w_{\alpha} \Big)\; ; \;\alpha = 1, \ldots, n_I\Big\} \, ,
\ee
where each dressed event, labeled by $\alpha$,  has fully specified jet substructure, \emph{i.e.},~specific values of $\vx_{1\,\alpha}$ and $\vx_{2\,\alpha}$ that have been drawn uniformly from their domain.  Since the $\vx_{i\,\alpha}$ are drawn uniformly (and not from the template itself), each dressed event is also associated with a corrected weight $w_\alpha^{\star}$ and an uncorrected weight $w_\alpha$:
\begin{eqnarray}
w_\alpha^{\star} = \srho(\vx_{1\,\alpha}\sep \vk_1)\;\srho(\vx_{2\,\alpha}\sep \vk_2) \qquad
\text{and} \qquad w_\alpha =  \trho(\vx_{1\,\alpha}\sep \vk_1)\;\trho(\vx_{2\,\alpha}\sep \vk_2). \label{eq:wStar}
\end{eqnarray}
The normalizations of $w^{\star}$ and $w$ are irrelevant, because they cancel when calculating efficiencies (see \eref{eq:Fprime}).  The corrected weight is used for the background prediction, while the uncorrected weight is used to estimate the total bias for the predicted number of events. 

To determine the final number of events that survive the cuts for a given search, one first computes the sum of corrected weights for each kinematic event before and after cuts: 
\be
W_i^\star = \sum_{\alpha =1}^{n_I} w^\star_{i\,\alpha} \qquad \text{and} \qquad
W_i^\star[C] = \sum_{\alpha =1}^{n_I} w^\star_{i\, \alpha}\; \Theta(c_{i\,\alpha}),
\ee
where
\begin{eqnarray}
c_{i\,\alpha} = \begin{cases}
1 & \text{ if } (\vx_1, \vx_2)_{i\,\alpha} \in C[\vx_1, \vx_2]\\
0 & \text{ otherwise} \, .
\end{cases}
\end{eqnarray}
The estimate for the efficiency that a given event $i$ passes the cuts is
\be
\hat\epsilon_i^{\star} \equiv \frac{W_i^\star[C] }{W_i^\star}.
\ee
Note that this is just the discrete version of \eref{eq:effContinuous}.  These efficiencies are then summed to obtain the final background prediction. The predicted number of background events after cuts, $\hat N^\star_e[C]$, is 
\be
\label{eq:Fprime}
\hat N^\star_e[C]=\sum_{i=1}^{N_e} {\hat\epsilon}^\star_i[C]\, .
\ee

The statistical uncertainty in the estimate $\hat N^{\star}_e[C]$ is challenging to determine because of the implicit dependence of $\hat N^{\star}_e[C]$ on the training sample.  This leads to important correlations between the cut efficiencies, ${\hat\epsilon}^\star[C]$, for different kinematic events.  A convenient technique for computing errors makes use of the bootstrap (see~\emph{e.g.},~\cite{efron79, scott, hall1995bootstrap, horowitz2001bootstrap}). An ensemble of datasets (resampled from the training sample) is used to calculate the corresponding set of templates.  This yields an ensemble of estimates for $\hat N^{\star}_e[C]$ whose variance is calculated directly.  While this procedure is computationally expensive, it results in a robust estimate of the statistical uncertainty, as will be demonstrated in App.~\ref{app:KernelSmoothing}.

In more detail, for an ensemble of resampled datasets, a corresponding set of templates $\big\{\srho_b\big \}$ is computed, with $b$ ranging from $1$ to $\nbs$.  These, in turn, are used to compute an ensemble of background predictions $\mathcal{N}^{\star}_e[C]=\big\{\hat N^{\star}_e[C]_1, ... , \hat N^{\star}_e[C]_{\nbs} \big \}$ using \eref{eq:Fprime}.  The \jargon{smoothing variance} is given by:
\be
\hat\sigma_V^2 =
\frac{1}{\nbs-1}\sum_{b=1}^{\nbs}\Big(\hat N^{\star}_{e}[C]_b- \left < \mathcal{N}^{\star}_{e}[C] \right >_{\text{\tiny{BS}}}\Big)^2 \,\qquad\!\! \text{with} \!\!\qquad
\left < \mathcal{N}^{\star}_{e}[C] \right >_{\text{\tiny{BS}}} =\frac{1}{\nbs}\sum_{b=1}^{\nbs} \hat N^{\star}_{e}[C]_b\,.
\label{eq:EventVar}
\ee
In practice, we find that $\nbs \sim \OO(100)$ is sufficient.  \eref{eq:EventVar} provides a good estimate of the statistical uncertainty in $\hat N^{\star}_e[C]$ provided that the bias is under control (see App.~\ref{app:KernelSmoothing} for more details). To make sure this is the case, we estimate the size of the residual bias $\hat\sigma_B$ inherent in the estimate $\hat N^{\star}_e[C]$ by comparing $\hat N^{\star}_e[C]$ to $\hat N_e[C]$, where the latter estimate uses the uncorrected weights:
\be
\hat\epsilon_i[C] = \frac{W_i[C] }{W_i} \qquad \text{and}\qquad \hat N_e[C] = \sum_{i=1}^{N_e} \hat\epsilon_i[C] ,
\ee
with
\be
W_i = \sum_{\alpha =1}^{n_I} w_{i\,\alpha}\qquad \text{and} \qquad
W_i [C] = \sum_{\alpha =1}^{n_I} w_{i\, \alpha}\; \Theta(c_{i\,\alpha}) \, .
\ee
The contribution of each kinematic event, $i$, to the bias is the difference between the two efficiencies $\hat\epsilon^\star_i[C]$ and $\hat\epsilon_i[C]$.  Note that this difference is a signed quantity, and the biases from different kinematic events may add (in)coherently.  A measure of the total bias $\hat\sigma_B$ is
\be
\label{eq:EventBias}
\hat\sigma_B = \left| \hat N_e[C]-\hat N_e^\star[C] \right|= \left| \sum_{i=1}^{N_e}\,\big(\,\hat\epsilon_i[C]-\hat\epsilon^\star_i[C]\big) \right|. 
\ee
The probability distribution function for the bias is unknown \emph{a priori} and is in general not Gaussian.  In order for $\hat\sigma_V$ to be a reliable estimate of the statistical uncertainty in $\hat N_e^\star[C] $, we must have $\hat\sigma_B \ll \hat\sigma_V$. In practice, we choose a kernel width such that $\hat\sigma_B$ is a factor of 2 to 5 less than $\hat\sigma_V$.  Because bias correction is done at the template-level, $\hat{\sigma}_B$ is an imprecise measure of the
residual bias in $\hat N^\star_e[C]$. Consequently, $\hat{\sigma}_B$ is taken to be unsigned and can be interpreted as a systematic error.

In addition to the statistical fluctuation of the templates, there is an additional systematic error $\sigma_\text{sys}$ coming from the mismatch between the template in the control region and the associated template that would be derived from a background-only sample in the signal region.  For the examples presented in the next section, we compute $\hat\sigma_V$ and $\hat\sigma_B$ explicitly.  These results demonstrate that $\sigma_\text{sys}$ is subdominant to the statistical errors associated with the template procedure, validating the approximation that the templates are independent.  

\subsection{Measurement Uncertainties}
\label{sec:measunc}
The preceding discussion made no mention of measurement uncertainties. If one were interested in comparing substructure templates extracted from data to analytical calculations, then the size and nature of any detector effects would be an important consideration in designing an appropriate template procedure. Indeed, the presence of these uncertainties introduces a number of complications (see \emph{e.g.},~\cite{Stefanski90deconvolutingkernel}).  In our case, templates are used to make background estimates that are then directly compared to data in the signal region.   Consequently, the substructure variables $\vx$ are implicitly taken to be the measured values; the templates are derived using the measured data and extrapolated to make a prediction for cuts on measured variables.  Furthermore, as long as detector-smearing is smooth, the kernel width can be taken to be smaller than the inherent detector resolution (and it will be in the limit of large statistics, see \eref{eq:SilvermanRule}).  This is not a problem, since what is being estimated is the convolution of the underlying kinematic distributions with the detector response.

The measurement uncertainties for the input variables $\vk$ are relevant if the kinematics are being modeled by MC.  In this situation, it is important that the kinematic sample be passed through a reliable detector simulator prior to the integration step so that all of the input and output variables correspond to quantities measured by the detector.  On the other hand, if the signal contamination in the region of interest is small (in the absence of any substructure cuts), it might be possible to take the kinematic sample directly from data, in which case these issues can be avoided.  

\section{Applications}
\label{sec:Applications}
This section is devoted to illustrating the data-driven template approach for two different searches.  The smoothing variance $\hat\sigma_V$ for 8 TeV LHC data is estimated and shown to be under control.  As we will see, the MC and template predictions are in excellent agreement. 

The first example is a search for high multiplicity signals (see Sec.~\ref{sec:HighMult}).  Fat jets from signal processes should exhibit accidental substructure~\cite{Hook:2012fd, Cohen:2012yc, Hedri:2013pvl} that would be suppressed for QCD backgrounds.  We demonstrate that the jet mass and $N$-subjettiness of the two leading jets of a four-jet sample can be predicted using templates derived from a three-jet sample.  For a second case study, the template method is applied to an existing ATLAS search for pair-produced gluinos undergoing an $R$-parity-violating (RPV) decay to three light quarks each (see Sec.~\ref{sec:RPVgluino}).  The observed correlation between the masses of the two leading jets is captured using the template approach.

Because we do not have access to actual LHC data, we rely on MC events.  Weighted events were generated using a variation on the binning procedure outlined in \cite{Avetisyan:2013onh}.  Specifically, two-, three-, and four-parton QCD events were generated using \texttt{Madgraph5}~\cite{Alwall:2011uj}.  These events were subsequently showered using \texttt{Pythia8}~\cite{Sjostrand:2007gs}; matching was done using the MLM procedure.  The particle-level events were grouped into cells to simulate the finite resolution of the calorimeter.  Trimming was applied to these pixels to reduce sensitivity to pileup effects (which we do not model), and $N$-subjettiness was computed using the \texttt{FastJet-contrib} plugin~\cite{Thaler:2010tr, Thaler:2011gf}.  For a full description of the simulation framework, see App.~\ref{sec:MCFramework}.

\subsection{High Multiplicity Signals}
\label{sec:HighMult}
This section demonstrates the application of template techniques to the types of high-multiplicity searches advocated in~\cite{Hook:2012fd, Cohen:2012yc, Hedri:2013pvl}.  This case study does not reproduce these searches in exact detail.  Instead, we study a simplified version of the event-subjettiness search in~\cite{Cohen:2012yc} in which the substructure of only the two leading jets is 
considered.  Despite the modification, this example is still complex enough to constitute a real test of the template methodology.  

As a preselection, each event is required to have at least $\nj=4$, $R=1.0$ anti-$k_T$ \cite{Cacciari:2008gp} jets with $\pT > 200 \GeV$. Cuts are placed on observables that depend on the fat jets' substructure (but not on their kinematics):  the sum of jet masses
\be
M_J = \sum_{i=1}^{2} m_i
\ee
and the `event-subjettiness'
\be
T_{21} =  \Bigg[ \prod_{i= 1}^2 \Big(\tau_{21}\Big)_{i} \Bigg]^{1/2} \, ,
\ee
where $\tau_{ij}$ is the $N$-subjettiness ratio $\tau_{j}/\tau_{i}$ \cite{Thaler:2010tr,Thaler:2011gf}.  Note that there is no impediment to replacing event-subjettiness by one of the subjet counting observables proposed in \cite{Hedri:2013pvl}, although the discrete nature of the latter would require modifications to the kernel smoothing procedure.

Requiring four fat jets is already extremely efficient at reducing the Standard Model background \cite{Hook:2012fd}.  An exclusive three-jet sample is therefore expected to be signal-poor, making it an ideal training sample.  The two leading jets in each event are used to fill a 3-dimensional template $\trho^{\star}(m, \tau_{21} \sep \,\pT)$.  

A good set of coordinates for the template is:
\be
\label{eq:MultijetCoords}
\trho^{\star} = \trho^{\star} \Bigg ( -\log_{10}\bigg(\frac{m}{\pT}\bigg),\, \tau_{21},\, \ln\bigg(\frac{\pT}{200\GeV}\bigg) \Bigg).
\ee
The log-transformed variables are appropriate given the logarithmic evolution of the strong coupling constant with respect to the fat jet $\pT$ and the form of the collinear singularity that governs the generation of mass in QCD jets.
\begin{figure}[h!]
\centering
\includegraphics[width=0.43\textwidth]{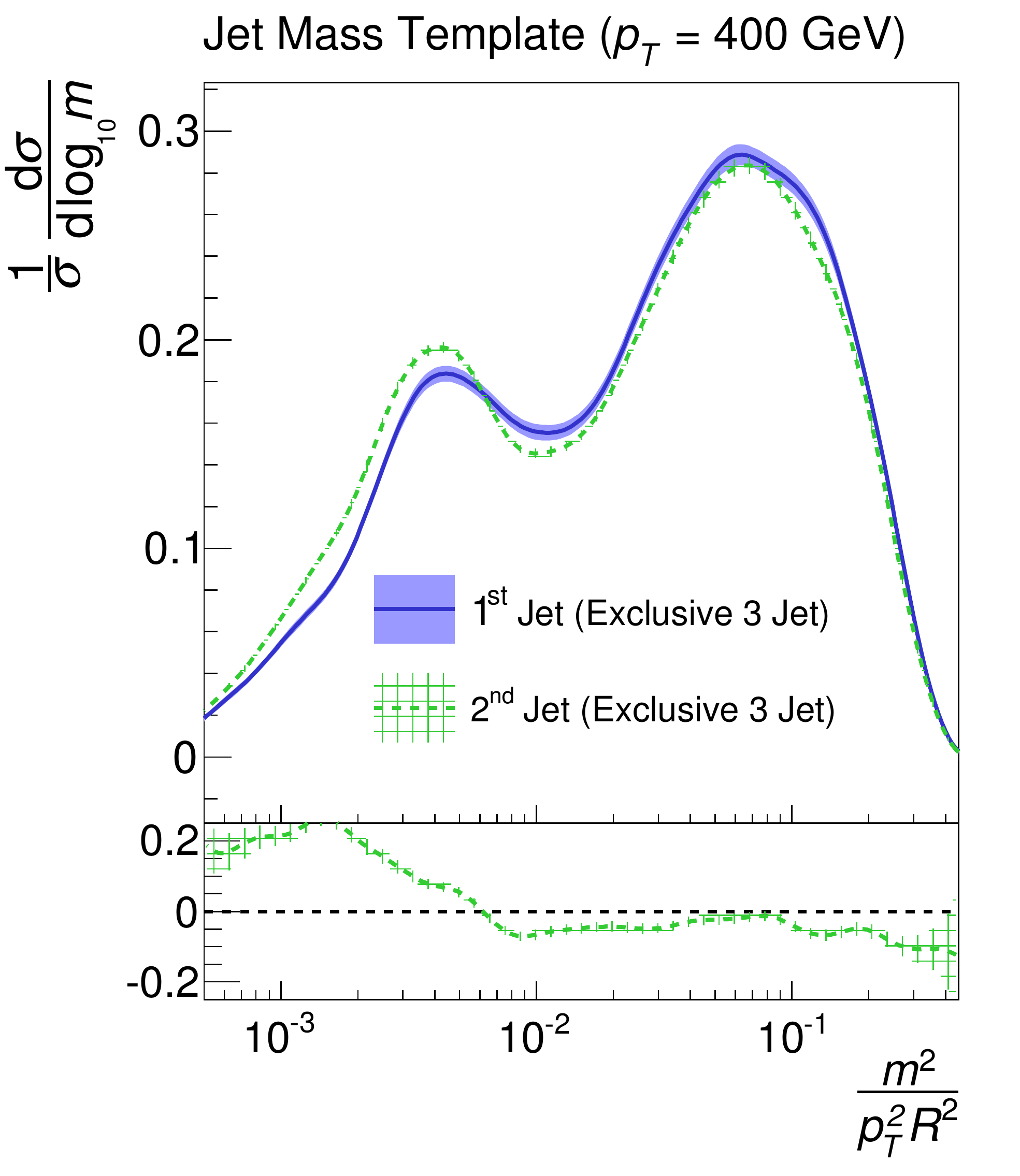} \includegraphics[width=0.43\textwidth]{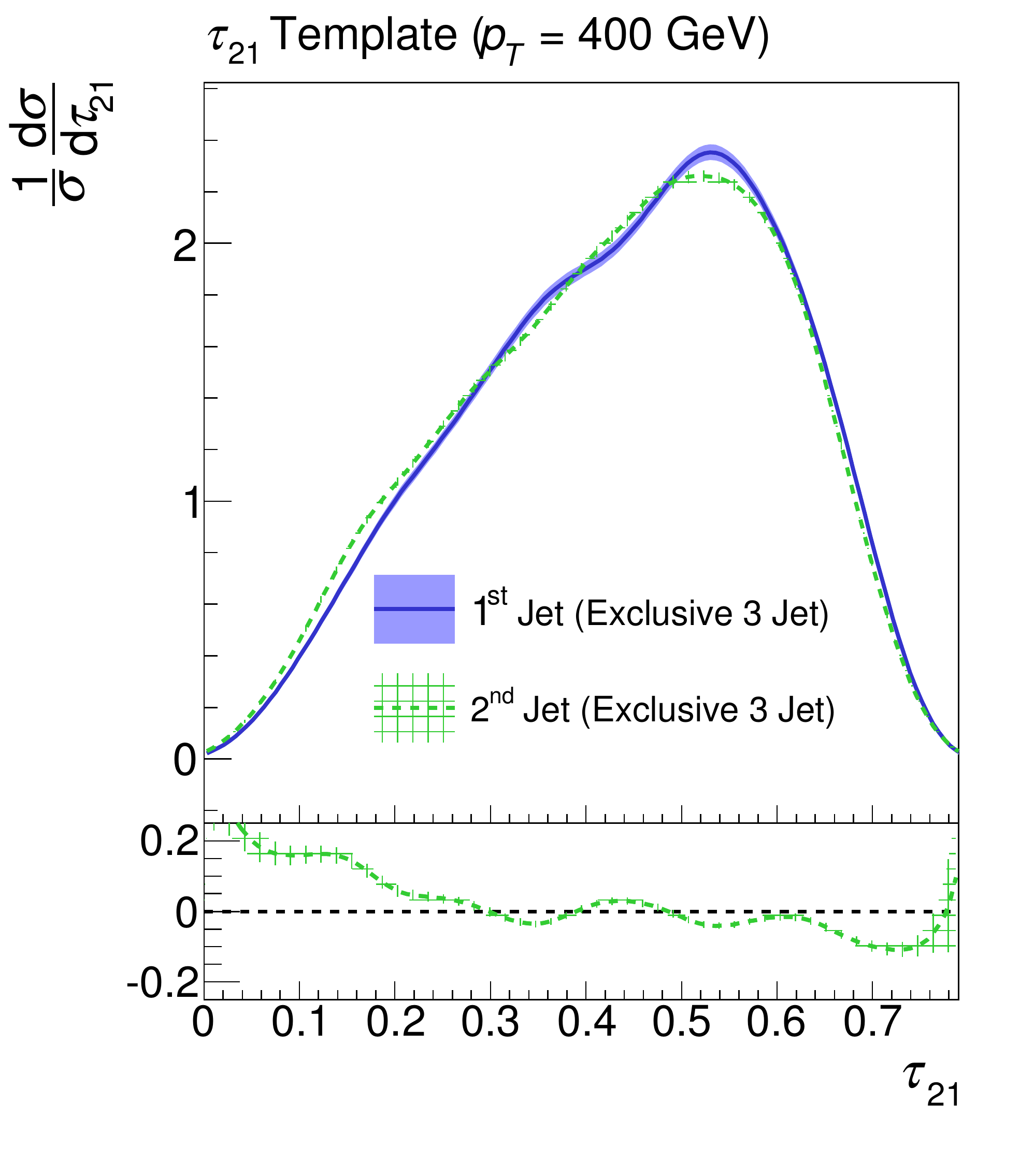}
\caption{
The plot on the left (right) compares the jet mass ($\tau_{21}$) templates between the first and second jets in the exclusive three-jet bin. The lower panels show the deviation between the two templates, normalized to the template of the leading jet. The error bands on the templates are statistical, calculated using the bootstrapped ensemble of templates.
}
\label{fig:template_diffjet}
\end{figure}

\begin{figure}[b!]
\centering
\includegraphics[width=0.43\textwidth]{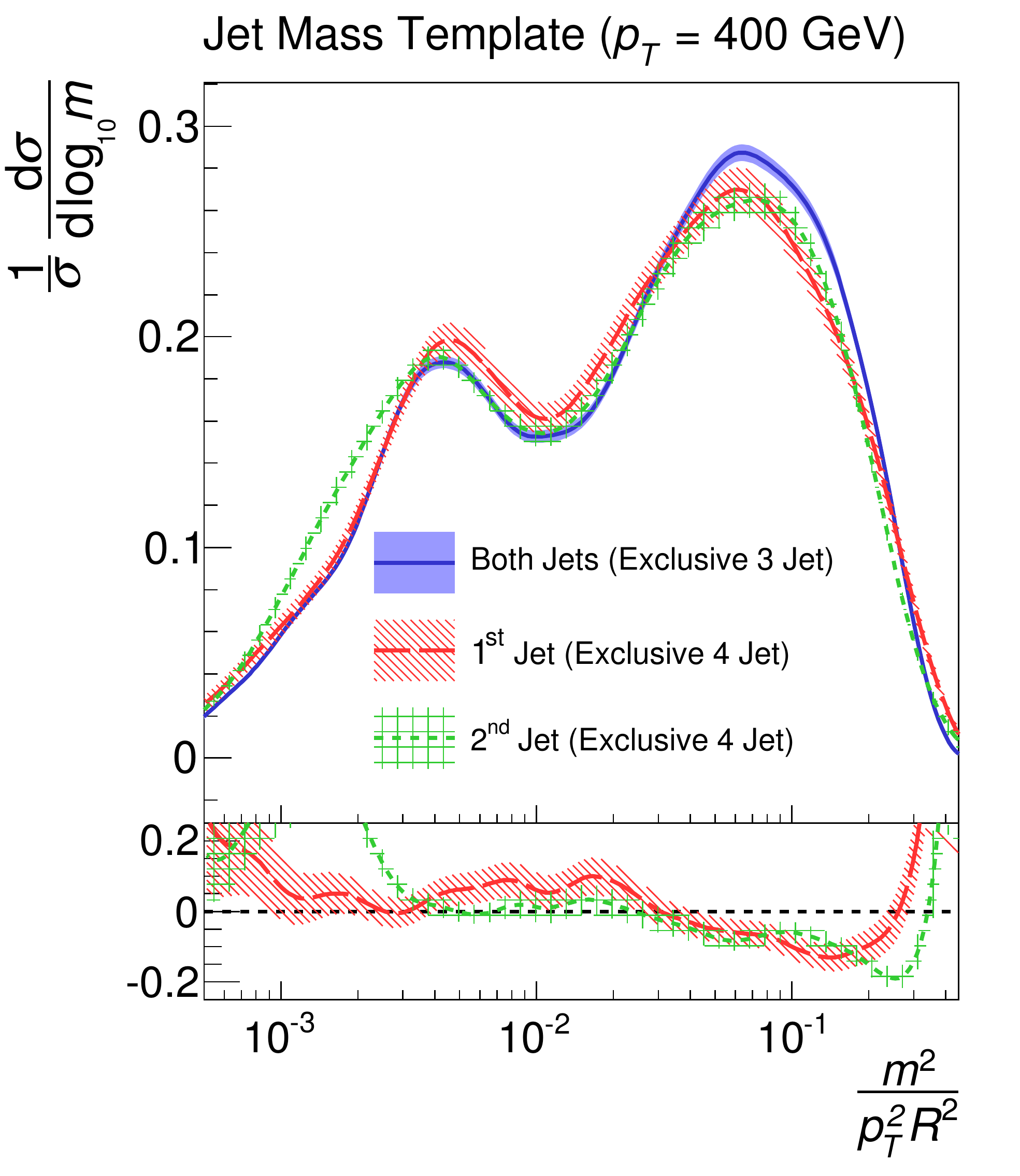} \includegraphics[width=0.43\textwidth]{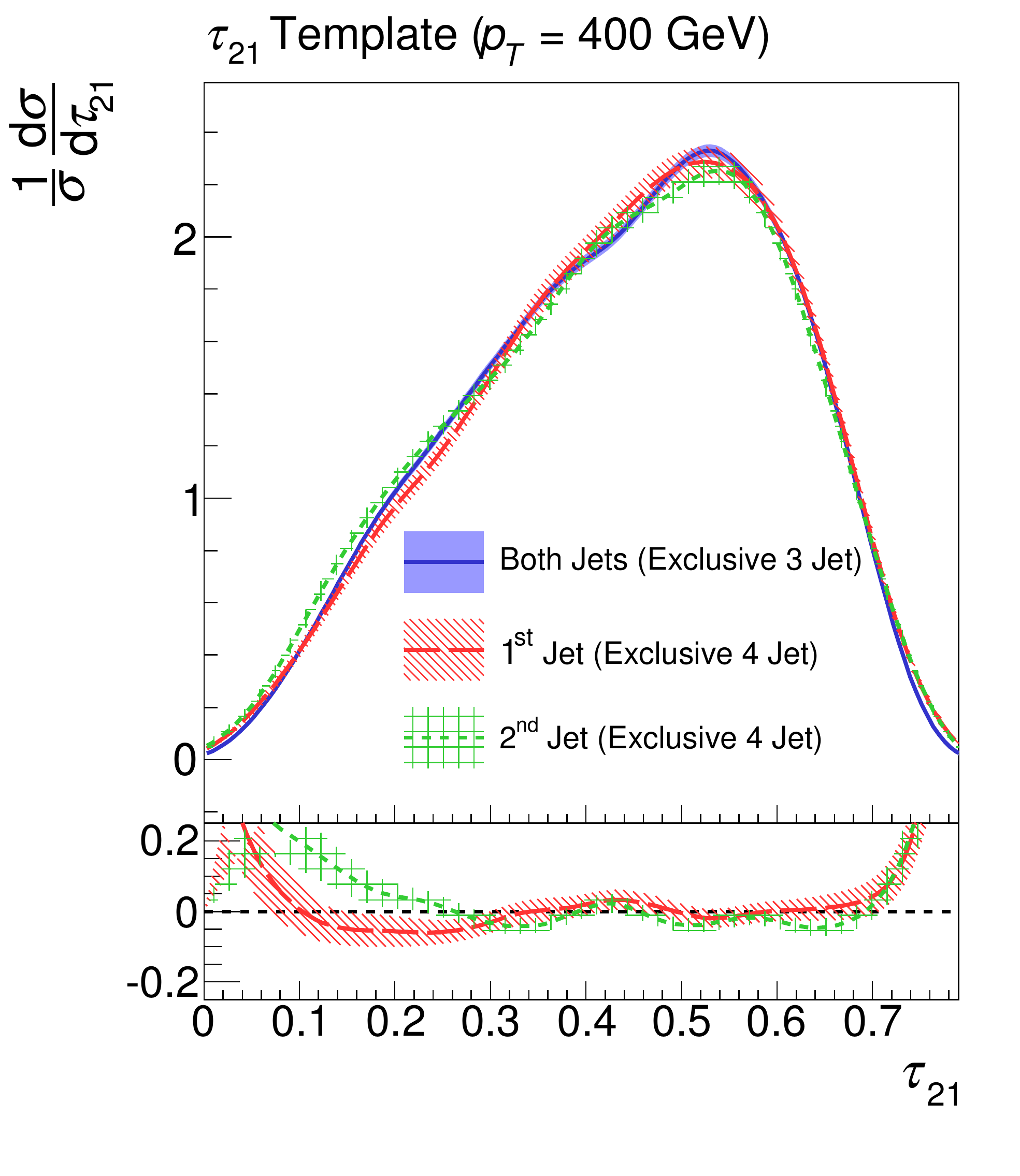}
\caption{
The plot on the left (right) compares the jet mass ($\tau_{21}$) templates between different exclusive jet bins. The exclusive three-jet template (blue) is obtained from the two leading jets. The error bands only include the template errors, computed using the bootstrapped ensemble of templates. The lower panels show the deviation between the templates, normalized to the three-jet template.
}
\label{fig:template_diffmulti}
\end{figure}

The bandwidth is chosen to produce a small total bias, $\hat\sigma_B \lesssim 
\hat\sigma_V$, while keeping the error $\hat\sigma_V$ as small as possible.  Fig.~\ref{fig:template_diffjet} shows the resulting templates, while Fig.~\ref{fig:template_diffmulti} compares the 3-jet template with the 4-jet prediction for a fixed $p_T$ slice.  These figures demonstrate that the substructure variables for the two leading jets in the exclusive 3-jet and 4-jet samples are equivalent within $10\%$ in the high-mass region and a broad range of $\tau_{21}$.  We have checked that this holds for a wide range of $p_T$ choices.

With the template in hand, the next step is to dress a kinematic sample, \emph{i.e.}, the integration step.  An inclusive 4-jet cut, with each $R=1.0$ anti-$k_T$ jet required to have $\pT >$ 200 GeV, fully characterizes the kinematic sample.  The leading and subleading jets (but not the $3^{\rm{rd}}$ or $4^{\rm{th}}$ jets) in the kinematic sample are then dressed with the substructure template.  The result is a dressed 4-jet sample in which the leading and subleading jets are associated with values of $\tau_{21}$ and fat jet mass $m_{1,2}$.  

\begin{figure}[h!]
\centering
\includegraphics[width=0.48\textwidth]{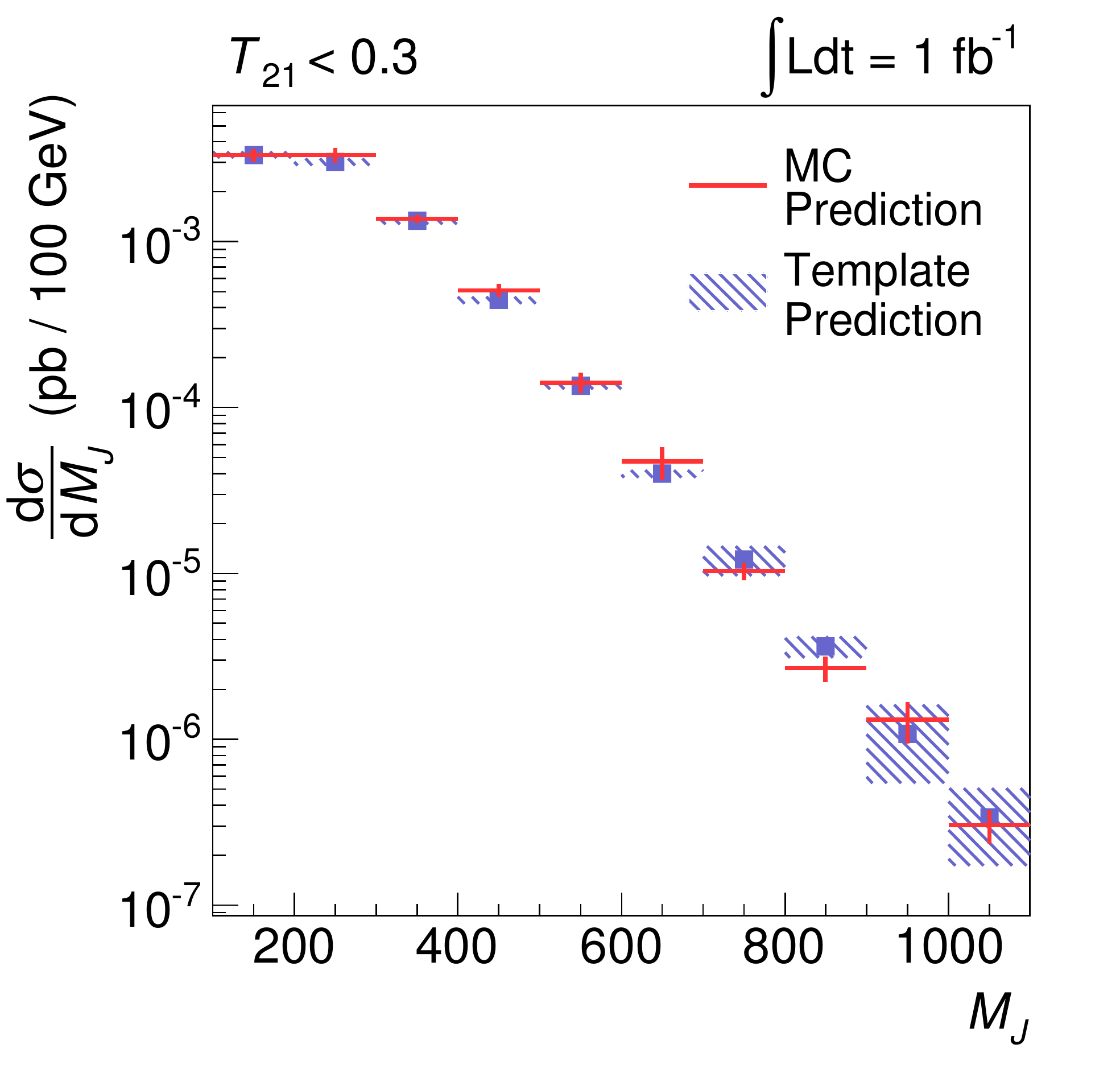} \includegraphics[width=0.48\textwidth]{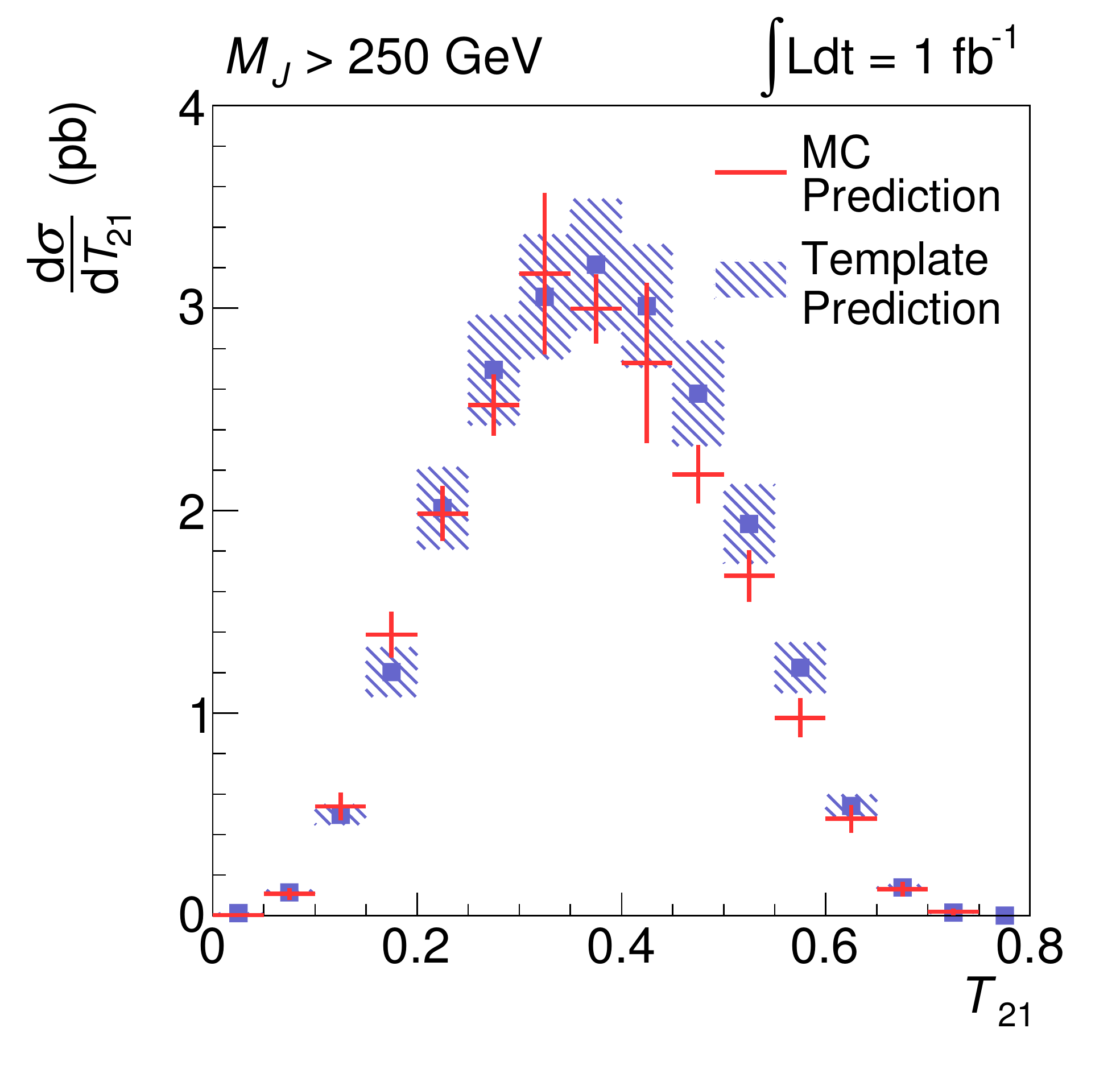}
\caption{
A comparison between the template estimate and the MC for the differential distributions for $M_{J}$ (with $T_{21}<0.3$) and $T_{21}$ (with $M_{J}>250$ GeV) for the high multiplicity case study.  The MC error bands show the statistical uncertainty for the weighted event sample.  The template error bands are given by $\hat\sigma_V$ in each bin (note that the errors are correlated). 
}
\label{fig:DressedPlot}
\end{figure}

Using the dressed sample, one can compute the fraction of dressed events passing the cuts.  We apply a preselection that \mbox{$m_j > 20$ GeV} for each jet, ensuring that the results are IR-safe  \cite{Larkoski:2013paa}.  Another preselection cut of \mbox{$p_T > 250$ GeV} is also applied to ensure insensitivity to the template boundary.  The resulting differential distributions for $M_{J}$ (with $T_{21}<0.3$) and $T_{21}$ (with $M_{J}>250$ GeV) are shown in Fig.~\ref{fig:DressedPlot}.  These distributions are in excellent agreement with the MC. Several additional cuts are summarized in Table \ref{table:multijet}; the template predictions agree with the MC to within template errors. 

\begin{table}[tb]
\renewcommand{\arraystretch}{1.4}
\setlength{\tabcolsep}{7pt}
\begin{tabular}{|c||c|c||c|c|}
\hline
$c$ &$M_J$ cut [GeV] & $T_{21}$ cut & MC & Template $\pm\, \hat\sigma_V \pm \hat\sigma_B$ \\
\hline \hline
0.37 & 500 & 0.3 &  20.3 $\pm$ 2.2 & 19.2 $\pm$ 2.3 $\pm$ 0.6 \\
0.52 & 750 & 0.3 & 0.86 $\pm$ 0.10 & 0.96 $\pm$ 0.19 $\pm$ 0.05 \\
0.37 & 500 & 0.6 & 45.8 $\pm$ 3.5 & 45.2 $\pm$ 3.7 $\pm$ 1.3 \\
0.52 & 750 & 0.6 & 1.67 $\pm$ 0.14 & 1.90 $\pm$ 0.19 $\pm$ 0.13 \\
\hline
\end{tabular}
\caption{Expected number of background events for the high multiplicity case study, assuming an integrated luminosity of $1\textrm{ fb}^{-1}$. 
The amount of smoothing used for each background estimate is specified in the first column (in units of $c_{\text{\tiny{AMISE}}}$). 
The errors for the template estimates are given by $\hat\sigma_V$, computed using \eref{eq:EventVar}.  The template and MC predictions agree to within the calculated errors.  $\hat\sigma_B$ is not normally distributed and therefore cannot be simply combined with $\hat\sigma_V$.}
\label{table:multijet}
\end{table}

Note that only the first and second jets in the kinematic sample were dressed in this case study.  Empirically, we find that the template derived from the two leading jets in the exclusive three-jet sample accurately models the two leading jets in the four-jet sample.  While the template for the third jet is qualitatively different, it does provide a good model of the third jet in the four-jet sample, although this statement suffers from large statistical uncertainties given the size of the MC sample.  The difference in the third-jet template is driven in large part by the average quark/gluon content of the third jet.  It is likely that by incorporating quark and gluon information (extracted from MC) into the determination of the templates, the procedure could be generalized to incorporate the third and even fourth jets into the analysis.  Exploring such a generalization is left for future work \cite{FutureWork}.  

\subsection{Boosted Three-Jet Resonances}
\label{sec:RPVgluino}
The second case study is based on the ATLAS search for boosted gluinos undergoing RPV decays to three light quarks \cite{ATLAS:2012dp}.  
The ATLAS analysis proceeds by clustering the event into fat $R=1.0$ anti-$k_T$ jets and narrow $R=0.4$ jets.  Several search regions are considered, each with a combination of cuts on narrow jet quantities (multiplicity, scalar sum $\pT$) and fat jet quantities (multiplicity, jet masses).  In addition, a substructure requirement of $\tau_{32} <0.7$ is imposed on both fat jets in order to select on the expected three-pronged structure of gluino decays.  The case study presented here mirrors the fat jet analysis, but ignores the narrow jet selections.  Note that we compute $N$-subjettiness using the ``min-axes" algorithm, whereas ATLAS used the ``$k_T$-axes" algorithm. 

To proceed with a background estimate as outlined in the previous two sections, we must first define a training sample from 
which we can construct a substructure template.  As a preselection on the MC sample, each event is required to have at least two fat jets with $\pT >$ 320 GeV.  Because the signal region consists of events with two high-mass fat jets, the training sample is defined by requiring at least one low mass jet with $m_j <$ 140 GeV.  This ensures that signal contamination in the training sample is small. That is, for every pair of leading and subleading fat jets $(j_1, j_2)$ the three-dimensional template $\trho^{\star}(m, \tau_{32} \sep\, \pT)$ is filled with $j_2$ whenever $m_{j_1} < 140$ GeV.  The procedure is then repeated for $(j_2, j_1)$.  
The template is formed using the coordinates 
\be
\trho^{\star} = \trho^{\star} \Bigg ( -\log_{10}\bigg(\frac{m}{\pT}\bigg), \tau_{32} \, , \ln\bigg(\frac{\pT}{320\GeV}\bigg)\Bigg),
\ee
as in the previous study.

The next step is to define a kinematic sample. In general, the natural way to define the kinematic sample is to use the same cuts as the signal region; apart from substructure cuts, the two samples are defined identically.  In the present case, this corresponds to choosing a sample generated by \texttt{MadGraph5} at $\sqrt{s}=8$ TeV, with each event required to have at least two fat jets each with $\pT >$ 350 GeV and no requirements on the fat jets' masses or values of $\tau_{32}$.  In the integration step, both of the fat jets in the kinematic sample are dressed with the template $\trho^\star(m, \tau_{32} \sep\, \pT)$.  This results in a dressed data set in which each event consists of a pair of fat jets, with given values of $\pT$, $m$, and $\tau_{32}$.  
\begin{table}[tb]
\renewcommand{\arraystretch}{1.4}
\setlength{\tabcolsep}{7pt}
\begin{tabular}{|c||c|c||c|c|}
\hline
c & $m_J$ cut [GeV] & $\tau_{32}$ cut & MC & Template $\pm\,\hat\sigma_V\pm\hat\sigma_B$ \\
\hline \hline
1.0 & 400 & 0.5 & 220 $\pm$ 26 & 210 $\pm$ 14 $\pm$ 6 \\
1.0 & 600 & 0.5 & 3.70 $\pm$ 0.32 & 4.70 $\pm$ 0.98 $\pm$ 0.05 \\
1.0 & 500 & 0.6 & 67.0 $\pm$ 3.4 & 73.0 $\pm$ 10.0 $\pm$ 2.6 \\
1.0 & 600 & 0.6 & 9.1 $\pm$ 0.7 & 11.0 $\pm$ 2.0 $\pm$ 0.3 \\
\hline
\end{tabular}
\caption{Expected number of background events for the RPV case study, assuming an integrated luminosity of $20\textrm{ fb}^{-1}$.  
The amount of smoothing used for each background estimate is specified in the first column (in units of $c_{\text{\tiny{AMISE}}}$). 
The errors for the template estimates are given by $\hat\sigma_V$, computed using \eref{eq:EventVar}.  The template and MC predictions agree to within the calculated errors.  $\hat\sigma_B$ is not normally distributed and therefore cannot be simply combined with $\hat\sigma_V$. 
}
\label{table:rpv}
\end{table}

\begin{figure}[h!]
\centering
\includegraphics[width=0.48\textwidth]{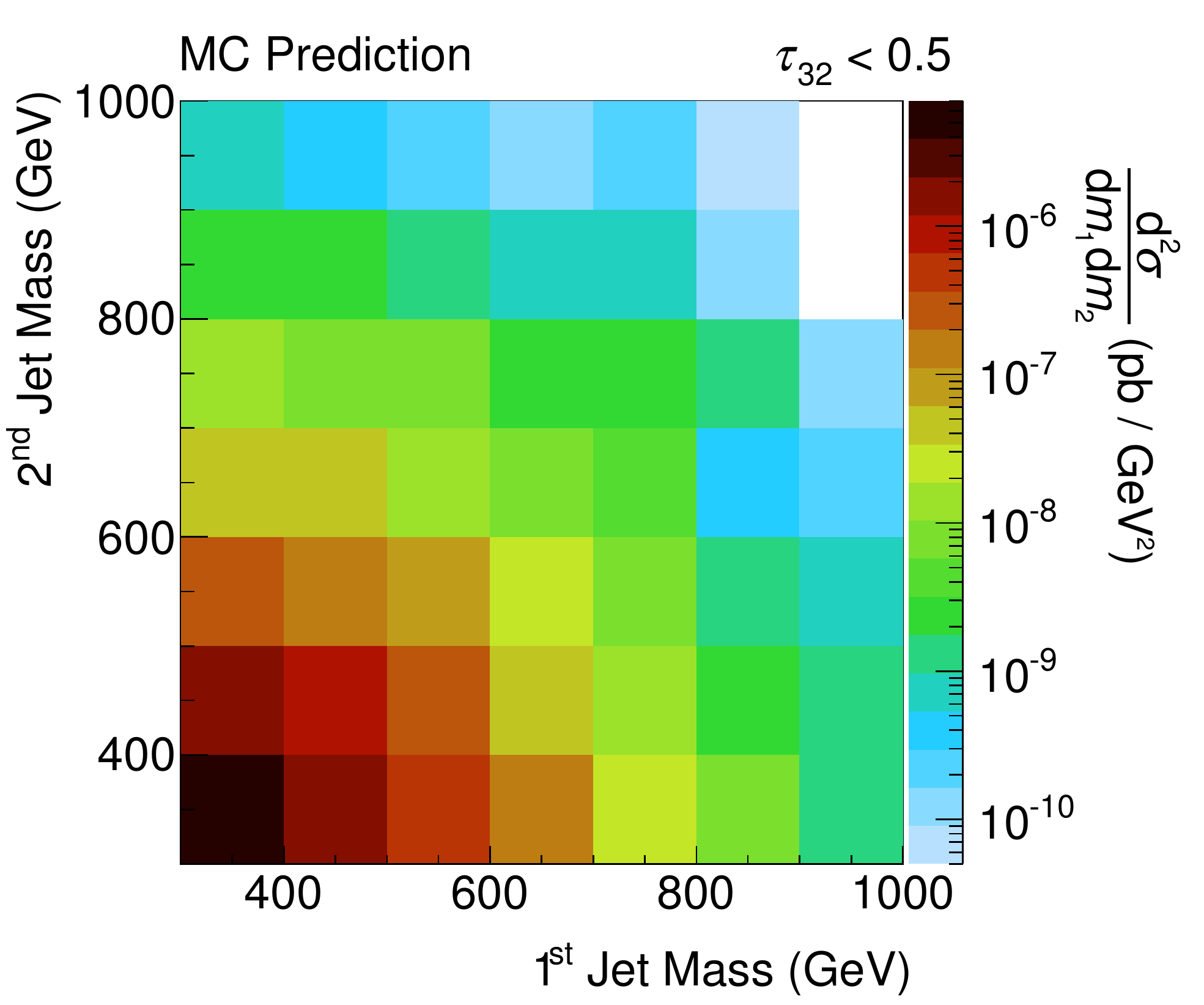} \includegraphics[width=0.48\textwidth]{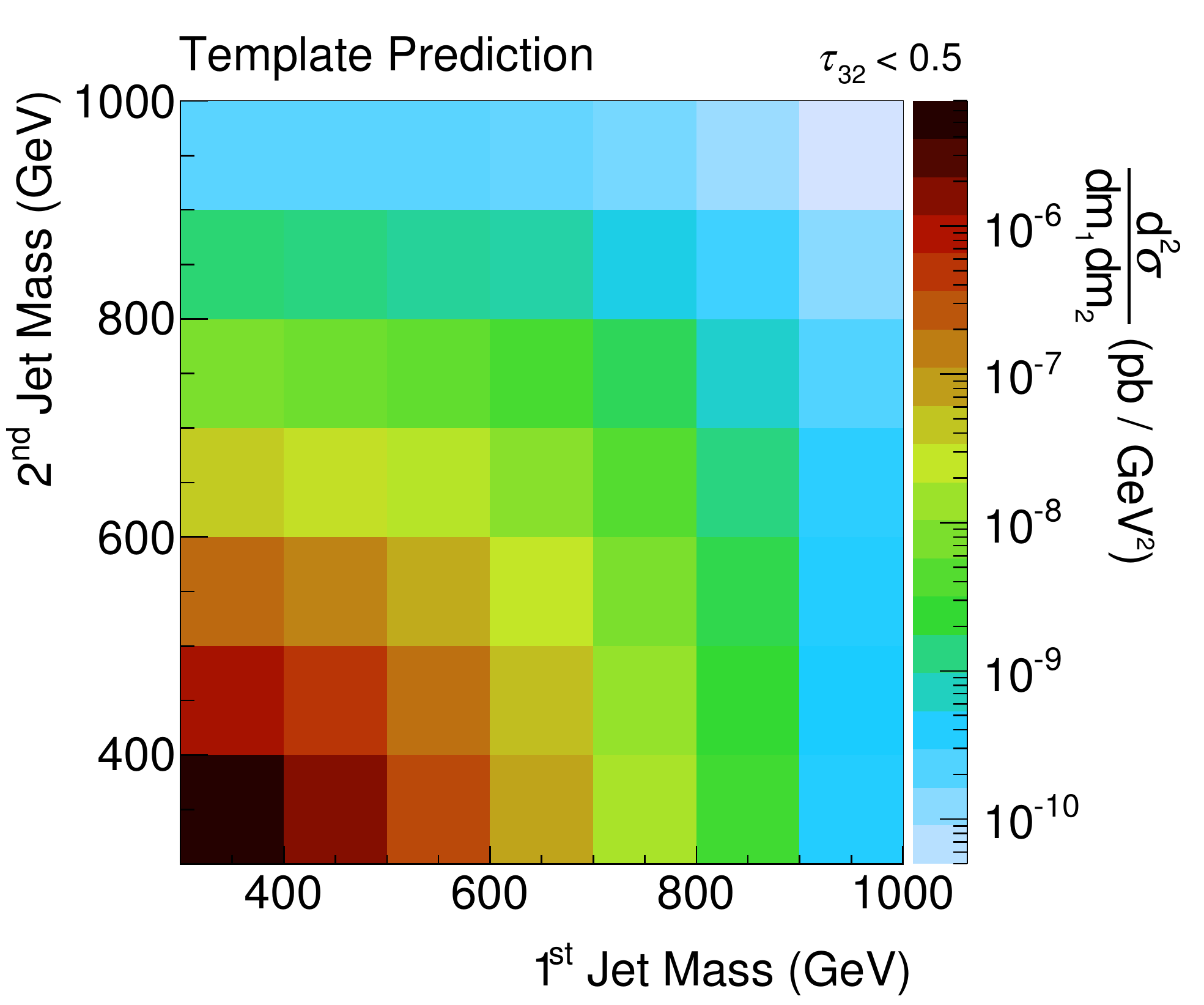}
\caption{The jet mass distributions of the two leading jets obtained directly from MC (left) and estimated using template methods (right).  The data-driven technique captures the dominant correlations between the jet masses.}
\label{fig:RPV_Mass}
\end{figure}

To assess the efficacy of the template approach, the dressed data set is compared to the full MC sample. One interesting test is to compare the correlations present in the $m_{j_1}$ vs.~$m_{j_2}$ plane in the kinematic sample with the results from the dressed data set. Fig.~\ref{fig:RPV_Mass} shows that the resulting jet mass distributions are indeed reproduced by the template estimate.  This figure also provides a good qualitative match with the relevant plot published by ATLAS \cite{ATLAS:2012dp}. The dominant correlations between the two jet masses are captured by virtue of the $p_{T}$ dependence of the $\trho^\star$ template. 

As in the previous study, the dressed data set can also be used to estimate yields in the signal region after imposing
substructure cuts. The predicted cross sections are listed in Table \ref{table:rpv} for a selection of cuts.  There is excellent agreement between the MC and template predictions.  One expects the smoothing variance at the LHC to be reduced when a large dataset is available for training the templates. Note that as the training sample increases, statistical uncertainties decrease more slowly than the familiar $1/\sqrt{N_T}$, see \eref{eq:SilvermanRule}.

\pagebreak
\section{Conclusions}
\label{sec:Conclusions}

Recent years have seen a revolution in fat jet techniques that could allow for the discovery of new physics in signal regions that would otherwise have remained obscured by QCD backgrounds.  Realizing the full potential of this statement assumes that we can reliably model the relevant backgrounds.  This strongly motivates developing new methods for data-driven background estimates, particularly for the challenging QCD background.

An attractive feature of the fat jet paradigm is that it offers a natural division of phase space into inputs (kinematics) and outputs (substructure variables) as a consequence of jet factorization.  This work takes advantage of this division to develop a novel method of estimating QCD backgrounds from data.  The method makes use of jet substructure templates that are obtained from a control region and used to dress a kinematic sample in the signal region.  The implementation chosen here depends on kernel smoothing techniques, which are used to obtain a statistical error for the background prediction.  For the two case studies in Sec.~\ref{sec:Applications}, the QCD MC was treated as mock data and the predictions from kernel smoothing were shown to match the MC predictions.  This illustrates that the systematic error introduced by treating the two leading jets as independent and identical is subdominant to the statistical error. 

It is worth contrasting our data-driven proposal with existing approaches, such as the ``ABCD" method.  Ideally, the ABCD method uses a highly uncorrelated pair of variables in order to define signal and control regions.  If correlations are non-negligible, MC is used to calculate a correction factor that is folded into the background extrapolation.  Thus we should consider two cases in comparing the two methods.  In the first case, uncorrelated variables can be found, and ABCD yields a data-driven background estimate without relying on MC.  Because the systematics associated with template methods and the ABCD method are complementary, each provides a valuable crosscheck on the background determination.  In the second case, uncorrelated variables cannot be found, and any ABCD estimate is strongly dependent on MC.  Template-based methods, however, may still be able to provide a data-driven estimate. This latter situation is expected to hold in many jet substructure searches of interest, \emph{e.g.}, for searches involving cuts on jet mass and additional substructure observables (because these observables are in general highly correlated).  

\begin{figure}[tb] 
   \centering
   \includegraphics[width=3in]{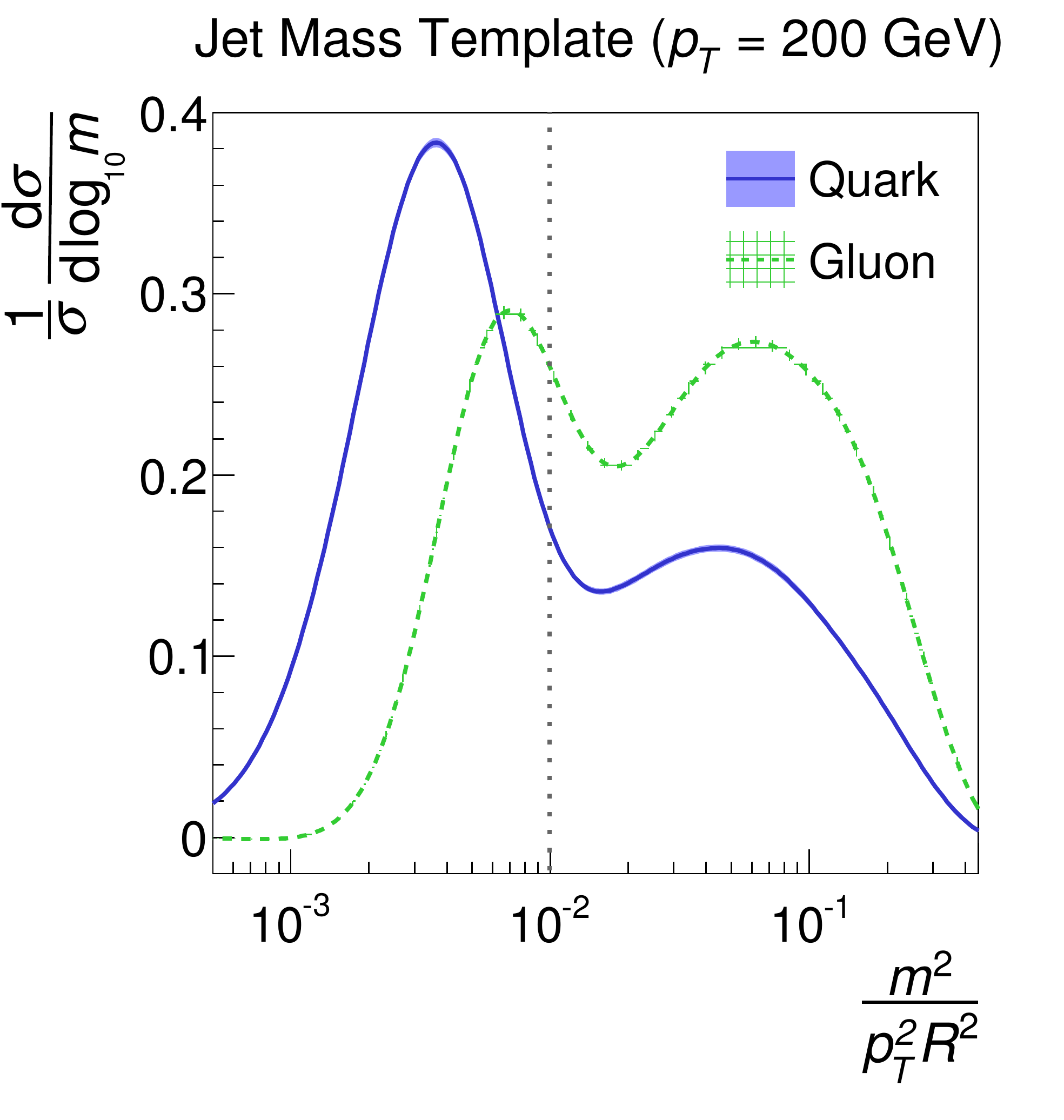} 
     \includegraphics[width=3in]{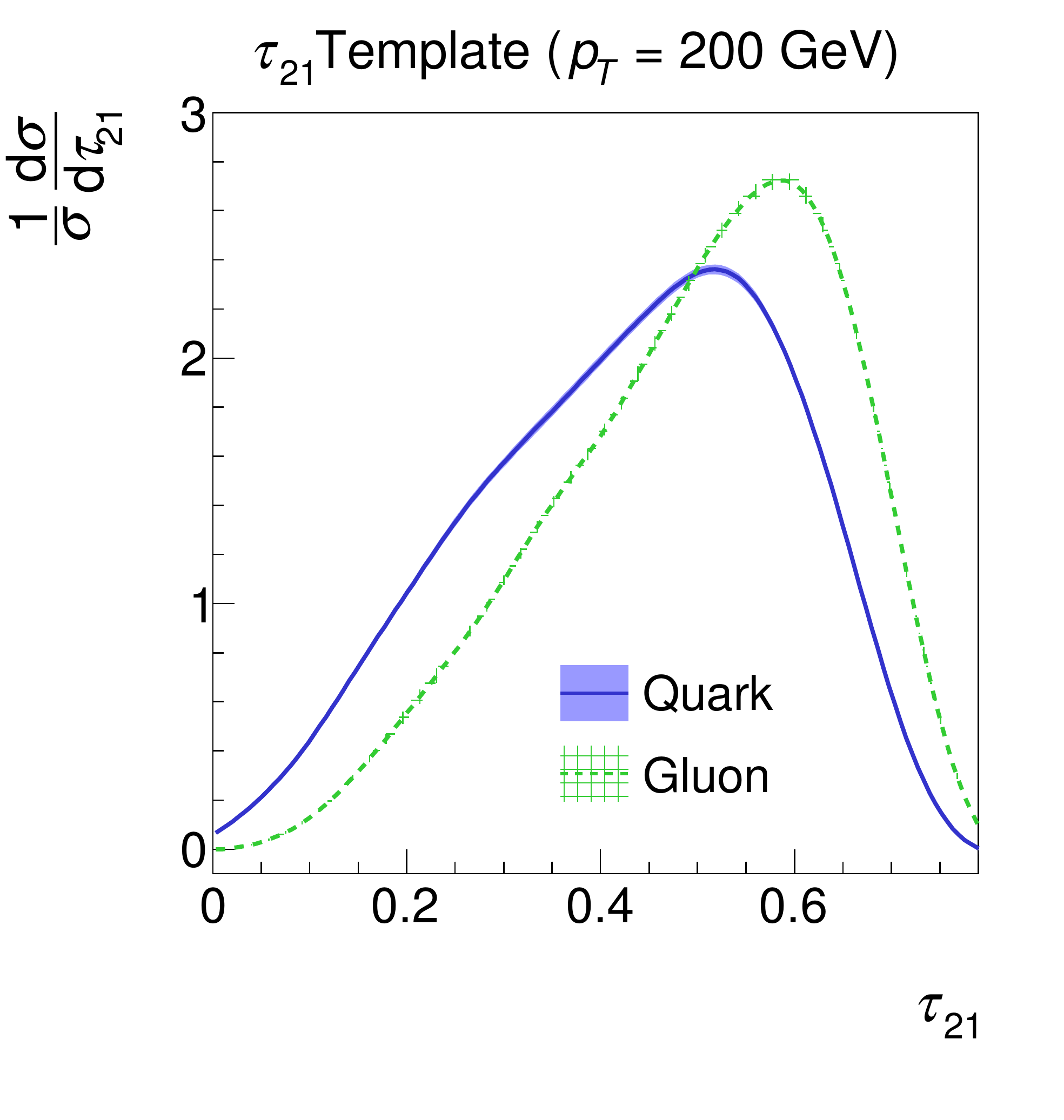} 
   \caption{Quark and gluon templates for jet mass (left) and $\tau_{21}$ (right).  A 20 GeV jet mass cut is indicated by the vertical grey line.}
   \label{fig:gluonquarktemplates}
\end{figure}

This paper is just the start of exploring applications of the template approach.  One important area of study is how templates depend on the quark/gluon composition of jets.  The techniques in this paper can be generalized to include separate templates for quarks and gluons, which can be viewed as a discrete input label.  To explore how the templates depend on quark/gluon content, we generated two MC samples of pure-quark and pure-gluon dijets and created the associated templates, shown in Fig.~\ref{fig:gluonquarktemplates}.  Once a \mbox{20 GeV} mass cut is imposed on each jet (dashed gray line), the quark and gluon mass distributions are very similar.  However, there is clearly information that could potentially be incorporated into the procedure.  Using separate quark/gluon templates will be a crucial step in applying these methods to third (or higher) jets in the event, which tend to be gluon jets.    

There are other directions that deserve further study.  It will be important to understand how the template estimates depend on the fat jet radius, and the impact of overlapping jets and pile-up.  While the jet factorization assumption facilitates straightforward data-driven estimates, it would be interesting to explore what happens when this condition is relaxed.  This might include systematically improving the template method with input from perturbative QCD calculations.  Finally, the template technique can be generalized to many different new physics searches, with different kinematic inputs and substructure variables.  

Template functions provide a powerful way to estimate backgrounds in regions where direct MC calculations are prohibitively hard.  This allows extrapolation of both MC and data into signal-rich regions.   A deviation would be the first sign of new physics or, at worst, an indication of the breakdown of jet factorization or novel QCD effects not included in MC.  No matter the result, we will learn something from the application of template methods to data.

\section*{Acknowledgements}
We thank G.~Salam, M.~Swiatlowski, and B. Tweedie for helpful discussions, as well as J.~Thaler for feedback on the draft.  TC and JGW are supported by DoE contract number DE-AC02-76SF00515.  HKL is supported by the DOE SCG Fellowship.  TC and ML thank the Galileo Galilei Institute and the INFN, the Aspen Center for Physics, and the NSF Grant No.~PHY-1066293 for partial support during the completion of this work.
\pagebreak

\appendix
\section*{Appendices}
\addcontentsline{toc}{section}{\hspace{23.5pt}Appendices}
\section{A Kernel Smoothing Primer}
\label{app:KernelSmoothing}
This appendix provides a detailed introduction to kernel smoothing and presents the key derivations that justify the procedures used in the main body of the paper.  It is divided into three parts: App.~\ref{sec:A_motivation} reviews the basic formulae and the main properties of kernel smoothing; App.~\ref{sec:A_implementation} details the various algorithmic choices that underlie our template procedure; and App.~\ref{sec:A_validation} validates the procedure on a test probability distribution.  We chose our particular prescription for its relative simplicity, computational efficiency, and the fact that it works well in MC. More sophisticated variations are possible, including for example, adaptive kernel smoothing (which uses a variable bandwidth) and boundary kernels (which improve performance near the edge of the domain).

A basic statistical problem is to estimate a probability distribution $\rho(\vec{z})$ from a random sample of finite size. The simplest strategy is to bin the data into a histogram. Histograms, however, are intrinsically non-smooth, and the resulting estimates can be very sensitive to the choice of binning scheme, especially when the sample size is small. Furthermore, the statistical errors associated with histograms can be difficult to compute. 

A better alternative is to use kernel smoothing techniques \cite{hastie,scott,Silverman86,Cranmer:2000du}.  Kernel smoothing replaces a histogram with a probability estimate $\trho(\vec{z})$, where $\vec{z}$ denotes a $D$-dimensional vector that can include kinematic and/or substructure variables. The weighting function, or kernel $K(\vec{z})$, must balance between the degree of locality and the smoothness of the resulting estimate.

Given a training sample with $N_T$ independent data points $\{\vec{z}_1, \vec{z}_2, ..., \vec{z}_{N_T}\}$ that have been obtained from the true distribution $\rho(\vec{z})$, one can obtain an estimate $\trho$ of the true density $\rho$ with the following master formula:
\be
\trho(\vec z)=\frac{1}{N_T}\sum_i^{N_T} K_h \big(\vec z-\vec z_i\big) \, ,
\qquad \text{where} \qquad
K_h(\vz_{\,}) = \frac{1}{\det h}K\big(h^{-1} \vz_{\,}\big) \, .
\label{eq:A_kernel}
\ee
Here, $K$ is the smoothing kernel (often chosen to be a Gaussian) and $h$ is a non-degenerate matrix that controls the smoothness of the resulting $\trho$ by changing the width of the kernel.  For a Gaussian kernel,\footnote{Because the Gaussian has infinite support, care should be taken when using this formula near (\emph{i.e.},~within a few bandwidths of) the boundary of the domain of $\vz$.}
\be
K(\vec z)=\frac{1}{(4\pi)^{\frac{D}{2}}}\exp
\left(-\frac{\vert\vec z\,\vert^2}{2}\right)
\quad \text{and} \quad
K_h(\vec z)=\frac{1}{(4\pi)^{\frac{D}{2}}\det h}\exp
\left[-\left(2h^Th\right)^{-1}_{\! ij}z^i z^j\right]\, .
\label{eq:A_kernel_Gaussian}
\ee
The density estimate is normalized, $\int \trho \,\,\dd^D{z} = 1$, as long as the kernel is normalized to unity, $\int K_h\, \dd^D{z} =1$. After fixing $K$ to be a Gaussian, one still has to choose the bandwidth matrix $h$, which should be large enough that the resulting estimate is reasonably smooth, but small enough that important features are not washed-out.  As we will see below, the choice of $h$ is typically more important than the choice of kernel, as it controls the size of the bias and variance.  Choosing the best value of $h$ is challenging because different measures result in different choices of $h$ (see \emph{e.g.},~\cite{Cao}).  The following subsections present a concrete prescription for choosing $h$ that leads to good statistical behavior.   

\subsection{Deriving the AMISE Bandwidth}
\label{sec:A_motivation}
To select an optimal bandwidth $h$, one must specify a metric by which to measure optimality.  A common choice is the Mean Integrated Squared Error (MISE), defined as the expectation of the square of the error term:

\begin{align}
\textrm{MISE} &=\int \dd^D z\; \left\langle \big(\hat{\rho}(\vz_{\,})-\rho(\vz_{\,}) \big)^2\right\rangle \\
&=
\int \dd^D z\; \big(\langle \hat{\rho}\rangle-\rho \big)^2 
+
\int \dd^D z\; \left\langle  (\hat{\rho}-\langle\hat\rho\rangle)^2 \right\rangle  \\
&\equiv \int \dd^D z\; b(\vz_{\,})^2 + \int \dd^D z\; v^2(\vz_{\,}) \, ,
\label{eq:A_MISE}
\end{align}
where the notation $\langle ... \rangle$ denotes the expectation value with respect to the true distribution; for any statistical quantity $f\big(\vec z_1, ..., \vec z_{N_T}\big)$ one defines
\be
\big\langle f(\vec z_1, \vec z_2, ..., \vec z_{N_T}) \big\rangle = \int \prod_i^{N_T} \dd^D{z}_i \, \,\rho(\vec z_i)\, f(\vec z_1, \vec z_2, ..., \vec z_{N_T})\,.
\ee
In \eref{eq:A_MISE}, the MISE is split into two terms.  The first is identified as the bias $b(\vz_{\,})$ squared and arises from the fact that kernel smoothing systematically smooths out peaks and valleys.  The second term yields the  variance $v^2(\vz_{\,})$, which encodes statistical fluctuations about the mean estimate $\langle \trho(\vz)\rangle$. \eref{eq:A_MISE} encapsulates the classic bias-variance trade-off in statistics: in order to reduce bias, one must reduce the amount of smoothing and hence increase the variance; to reduce the variance, one has to smooth out more features and hence increase the bias. 

Computing the bias and variance is difficult because each depends explicitly on the true distribution $\rho$ one is trying to estimate. Given a large sample size, it usually suffices to consider leading-order expressions in the asymptotic limit of large $N_T$. As the number of data points approaches infinity, the optimal bandwidth will approach zero, $h^{\text{\tiny{AMISE}}}\rightarrow 0$.  As a result, it is useful to expand the MISE around the limit $N_T\rightarrow \infty$ and $h \rightarrow 0$ to derive an asymptotic expression for the optimal bandwidth, $h^{\text{\tiny{AMISE}}}$.  Consider the average estimate $\langle \hat\rho (\vec z)\rangle$. In the large $N_T$ limit, the summation in \eref{eq:A_kernel} becomes a continuum integral:
\be
\Big\langle\trho\big(\vz_{\,}\big)\Big\rangle\rightarrow \frac{1}{\det h}\int \dd^D{z}'\,K\left(h^{-1}(\vec z-\vec{z}\,')\right)\rho\big(\vec{z}\,'\big) \, .
\ee
Introducing $\vec{s} = h^{-1}\big(\vec z-\vec{z}\,'\big)$, one has $\vec{z}\,' = \vec z-h\,\vec s$ and $\dd^D{z}\,' = \det h\,\dd^D s$ so that 
\be
\Big\langle\trho\big(\vz_{\,}\big)\Big\rangle = \int \dd^D s\,K\left(\vec s \right)\rho\big(\vec z-h\,\vec s_{\,} \big)\, .
\label{eq:A_rho_expectation}
\ee
Expanding the integrand to second order in $h$ and writing out all the indices explicitly yields
\bea
\Big\langle\trho\big(\vz_{\,}\big)\Big\rangle &\simeq& \int\dd^D s \left(  K\big(\vec s_{\,}\big) \,\rho\big(\vz_{\,}\big) - K\big(\vec s_{\,}\big)\,(h\,s)^i \frac{\dd\rho\big(\vz_{\,}\big)}{\dd z^i} + \frac{1}{2} K\big(\vec s_{\,}\big)\, (h\,s)^i (h\,s)^j \frac{\dd^2\rho\big(\vz_{\,}\big)}{\dd z_i \,\dd z_j} \right) \notag \\
&=& \rho\big(\vz_{\,}\big) + \frac{h_{ik}\,h_{jl}}{2}\, \frac{\dd^2\rho\big(\vz_{\,}\big)}{\dd x_i\, \dd x_j} \int \dd^D s \,s_k \,s_l \,K\big(\vec s_{\,}\big),
\eea
where a symmetric kernel with $K(s) = K(-s)$ has been assumed.  For the Gaussian kernel, which
satisfies $\int \dd^D s\,  s_i\, s_j\,K\big(\vec s_{\,}\big)= \delta_{ij}$, the leading-order bias is 
\be
b\big(\vz_{\,}\big)\equiv \Big \langle \trho\big(\vz_{\,}\big) \Big \rangle - \rho\big(\vz_{\,}\big) = \frac{(h^T h)_{ij}}{2} \frac{\dd^2\rho\big(\vz_{\,}\big)}{\dd z_i\, \dd z_j} \, .
\label{eq:TemplateBias}
\ee
The interpretation is that the smoothing bias is most pronounced at peaks and valleys where the second derivative is large.  Local maxima (minima) are estimated to be lower (higher) due to smoothing. 

A similar calculation can be performed for the variance:
\bea
v^2\big(\vz_{\,}\big)&=&\Big\langle\trho^2\big(\vz_{\,}\big) \Big\rangle- \Big\langle\trho\big(\vz_{\,}\big) \Big\rangle^2 \\
 &=&
\frac{1}{N_T^2}\left[ \int \prod_k^{N_T} \dd^D z_k \, \rho\big(\vz_{k}\big) \, \sum_{i,j}^{N_T} K_h\big(\vz-\vz_i\big)K_h\big(\vz-\vz_j\big) \right. - \notag \\
&\quad& \qquad \qquad \qquad 
\left. \left(\int \prod_k^{N_T} \dd^D\, z_k\, \rho\big(\vz_{k}\big) \sum_{i}^{N_T} K_h\big(\vz-\vz_i\big) \right)^{2\,\,} \right] \, .
\eea
In the first summation, the $i=j$ terms yield $N_T$ integrals over $K^2_h$. For $i\neq j$, each of the $N_T^2-N_T$ integrals factorizes into the square of $\int \dd^D z\,' K_h\big(\vz-\vz\,'\big)$, as $\vec z_i$ and $\vec z_j$ are independent random variables drawn from the same distribution.  In the second summation, independence of the $z_i$ yields a further $N_T^2$ terms $\big(\int \dd^D z\,' K_h\big(\vec z-\vec z\,'\big)\big)^2$. Gathering everything together, one has

\be
v^2(\vec z) = 
\frac{1}{N_T^2}\left[ N_T \int \dd^D z\,' \rho\big(\vz\,'\big) \, K^2_h\big(\vz-\vz\,'\big)
-N_T \left( \int \dd^D z'\, \rho\big(\vz\,'\big) K_h\big(\vz-\vz\,'\big) \right)^2 \right] .
\ee

Until this point, the computation is exact. To proceed further, consider the limit $h\rightarrow 0$, where $K_h\big(\vz_{\,}\big) \rightarrow \delta\big(\vz_{\,}\big)$. The first term dominates because it contains a factor of $\delta^2\big(\vz_{\,}\big)$.  Therefore, in the asymptotic limit:
\bea
v^2\big(\vz_{\,}\big) &\simeq& \frac{1}{N_T} \int \dd^D z' \rho\big(\vz\,'\big) \, K^2_h\big(\vz-\vz\,'\big) \\
&=&
\frac{1}{N_T\,\det h^2} \int \dd^D z' \rho\big(\vz\,'\big) \, K^2\big(h^{-1}\big(\vz-\vz\,'\big)\big) \\
&=& 
\frac{1}{N_T\,\det h} \int \dd^D s\,\, \rho\big(\vz-h\,\vec s_{\,}\big) \, K^2(s) \simeq \frac{\rho\big(\vz\big)}{N_T\,\det h} \int \dd^D s \, K^2(s),
\eea
where $\rho\big(\vz-h\,\vec s_{\,}\big)\simeq \rho\big(\vz_{\,}\big)$ to leading order in the last line.  The variance is directly proportional to the true density in the neighborhood of $\vec{z}$. For a Gaussian kernel, the integral evaluates to $\int \dd^D s \, K^2(s)=(4\pi)^{-\frac{D}{2}}$.  The leading-order variance in the asymptotic limit is then
\be
v^2\big(\vz_{\,}\big) = \frac{\rho\big(\vz_{\,}\big)}{N_T\,(4\pi)^{\frac{D}{2}}\det h} \,.
\label{eq:TemplateVar}
\ee

Armed with asymptotic formulae for the bias and variance, we can compute the Asymptotic MISE (AMISE) by plugging \eref{eq:TemplateBias} and \eref{eq:TemplateVar} into \eref{eq:A_MISE}:
\be
\textrm{AMISE} \equiv \int \dd^D z\, \left[ \left( \frac{(h^T h)_{ij}}{2} \frac{\dd^2\rho(\vec z)}{\dd z_i \,\dd z_j}\right)^2 +\frac{\rho\big(\vz_{\,}\big)}{N_T\, (4\pi)^{\frac{D}{2}}\,\det h} \right].
\label{eq:A_AMISE}
\ee
The optimal bandwidth is determined by minimizing the AMISE; this derivation yields a bandwidth $h^\text{\tiny{AMISE}}$ that is correct to leading order in $N_T^{-1}$. To explicitly minimize this expression, it is useful to rewrite the bias. Using integration by parts, \eref{eq:TemplateBias} is rewritten as
\be
\frac{1}{4} \big(h^T h\big)_{ij}\,\big(h^T h\big)_{kl} \int \dd^D z\, \frac{\rho\big(\vz_{\,}\big)\, \dd^4\rho\big(\vz_{\,}\big)}{\dd z_i\, \dd z_j\, \dd z_k \, \dd z_l}\equiv\frac{1}{4} \big(h^T h\big)_{ij}\,\big(h^T h\big)_{kl}\, S_{ijkl}\,,
\ee
where $S_{ijkl}$ is a totally symmetric rank-4 tensor.  Any such tensor can be diagonalized by an orthogonal transformation; it is possible to perform a coordinate transformation $\vec z \rightarrow \vec y$ such that $S_{ijkl}=\frac{1}{4}(\delta_{ij}\delta_{kl}+\delta_{ik}\delta_{jl}+\delta_{il}\delta_{jk})(4\pi)^{-\frac{D}{2}}$, where the normalization factor assumes $\rho$ has been transformed into standard normal form.  Defining $H_{ij}\equiv (h^T h)_{kl}\,\frac{\dd y_i}{\dd z_k}\frac{\dd y_j}{\dd z_l}$, the AMISE can be rewritten as
\be
\label{eq:AMISE_Final}
\textrm{AMISE} =  \frac{1}{(4\pi)^{\frac{D}{2}}\det \frac{\dd \vec y}{\dd \vec z}}\left(\frac{1}{16}\left[ 2\,\textrm{tr}\,H^2+(\textrm{tr}\,H)^2 \right] +\frac{1}{N_T \, \sqrt{\det H}}\right) \, .
\ee
The minimum occurs when $H_{ij}$ is proportional to the unit matrix $\delta_{ij}$. The optimal bandwidth in the asymptotic limit is then identified as\footnote{$h_{ij}$ is ambiguous up to an orthogonal transformation $h_{ij}\rightarrow O_{ik}\,h_{kj}$. This ambiguity is unimportant, as only $h^T h$ is relevant for Gaussian kernels.} 
\be
H^{\text{\tiny{AMISE}}}_{ij}=\delta_{ij}\left(\frac{4}{2+D}\right)^{\frac{2}{D+4}}N_T^{-\frac{2}{D+4}} \qquad \text{and} \qquad h^{\text{\tiny{AMISE}}}_{ij}= \frac{\dd y_i}{\dd z_j}\left(\frac{4}{2+D}\right)^{\frac{1}{D+4}}N_T^{-\frac{1}{D+4}} \, .
\label{eq:A_optimal_bandwidth}
\ee

Even in the asymptotic limit, the expression for the optimal bandwidth depends on the true distribution $\rho\big(\vz_{\,}\big)$ via $\frac{\dd y_i}{\dd z_j}$. This dependence cannot be removed without making further assumptions. For a Gaussian distribution, the required transformation is simply $y_i=\sigma_{ij}\,z_j$, where $\sigma_{ij}$ is the square root of the covariance matrix. For more general distributions that are not too far from the Gaussian distribution, one expects the correct transformation $z\rightarrow y$ to roughly correspond to rotating away the correlations and scaling the width of the distribution to unity.  Hence, the following choice of the bandwidth should be near optimal for most reasonable probability distributions:
\be
h^{\text{\tiny{AMISE}}}_{ij}\simeq c \,\sigma_{ij}\, N_T^{-\frac{1}{D+4}} \,.
\label{eq:A_silverman}
\ee
Here, $\sigma_{ij}$ is the square root of the covariance matrix computed from the distribution $\rho(\vec z)$, and $c$ is an $N_T$-independent $\OO(1)$ constant; \eref{eq:A_silverman} is known as Silverman's Rule-of-Thumb \cite{Silverman86}. In practice, $\sigma_{ij}$ is replaced by an estimate $\hat\sigma_{ij}$ from data.  As we will see in App.~\ref{sec:A_validation}, variance estimation is more reliable than bias estimation in practice.  It is therefore advantageous to choose a value for $c$ that deliberately undersmooths $\trho$, so that the bias is smaller than the variance and the error estimate can be trusted.

\subsection{Practical Kernel Smoothing}
\label{sec:A_implementation}
This subsection presents the specific algorithms and formulae used in this paper.  In the applications we consider, kernel estimates $\hat\rho\big(\vz_{\,}\big)$ are typically constructed from very large samples, and it is not always feasible to keep the $\vec{z}_i$ for each event in memory.\footnote{All events need to be stored in memory to implement the Fast Fourier Transform \cite{CooleyTukey,FFTW05} used in our \texttt{C++} code.}  To limit memory usage, the dataset is first binned with a small bin size $h_b \ll h$, and then kernel smoothing is performed on the binned dataset.  Additionally, to save computational time, the kernel estimate is precomputed at each bin location and linear interpolation is used to compute $\hat\rho\big(\vz_{\,}\big)$ for intermediate values of $\vec z$. 

More explicitly, let $\vec q$ denote the center of a given bin and let $\vec q\,\big(\vz_{\,}\big)$ be the function that sends $\vec z$ to the center of the corresponding bin.  Given a training sample $\{ \vec{z}_i \}$ of size $N_T$, denote the number of $\vec{z}_i$ in each bin $\vec q$ as $n_{\vec q}$. The master formula for kernel smoothing, \eref{eq:A_kernel}, then becomes
\be
\label{eq:binnedRhoHat}
\hat\rho\big(\vec q\,' \big)=\frac{\kappa}{N_T}\sum_{i}^{N_T} K_h\Big(\vec q\,'-\vec{q}\,(\vec{z}_i)\Big)=\frac{\kappa}{N_T}\sum_{\vec q} n_{\vec q}\, K_h\big(\vec q\,'-\vec q\,\big) \, ,
\ee
where $\kappa$ is a normalization constant. \eref{eq:binnedRhoHat} only defines $\hat\rho$ at the center of each bin; to extend the domain of $\hat\rho$ to all values of $\vec z$, multi-dimensional linear interpolation is performed.  In principle, the constant $\kappa$ can be set to normalize $\hat\rho\big(\vz_{\,} \big)$ to unity; when computing cut efficiencies this constant drops out, so it can be ignored in practice.

Combined with the rule-of-thumb for bandwidth selection given in \eref{eq:A_silverman}, the calculation of $\hat\rho\big(\vz_{\,} \big)$ is now straightforward.  Estimating the bias $\hat{b}\big(\vz_{\,} \big)$ and variance $\hat{v}^2\big(\vz_{\,} \big)$ is less trivial.
One could in principle substitute $\trho\big(\vz_{\,} \big)$ for $\rho\big(\vz_{\,} \big)$ into \eref{eq:TemplateBias} and \eref{eq:TemplateVar} and use the resulting estimates, which are correct to leading order in $h$.  We find, however, that it is preferable to use a different approach, which we now describe.\footnote{For example, because \eref{eq:TemplateBias} requires an estimation of the second derivative of $\rho$, it involves the additional complication of selecting a bandwidth appropriate for derivative estimation.}

The bias can be estimated by performing kernel smoothing twice and comparing the result with that obtained from performing kernel smoothing once. Specifically,\footnote{In principle, \eref{eq:TemplateBiasImp} depends on two distinct bandwidths.  In practice, we find that it is sufficient to take these to be equal as determined using \eref{eq:A_silverman}. It should be kept in mind, however, that different bias estimators are possible and may perform better on real data.}
\bea
\hat{b}\big(\vz_{\,} \big)&=&\int \dd^D z\,' \, \hat\rho\big(\vz-\vz\,'\big) K_{h}\big(\vz\,'\big) - \hat\rho\big(\vz_{\,} \big)
\label{eq:TemplateBiasImp}
\\
&\simeq& \frac{(h^T h)_{ij}}{2} \frac{\dd^2\trho\big(\vz_{\,} \big)}{\dd z_i \, \dd z_j}\simeq \frac{(h^T h)_{ij}}{2} \frac{\dd^2\rho\big(\vz_{\,} \big)}{\dd z_i\, \dd z_j} \notag \, .
\eea
This procedure is equivalent to \eref{eq:TemplateBias} in the asymptotic limit \cite{scott}.  We find that \eref{eq:TemplateBiasImp} yields relatively stable results in the test simulation discussed in App.~\ref{sec:A_validation}.   Note that the kernel smoothing in \eref{eq:TemplateBiasImp} is performed using the same algorithm outlined above.{\interfootnotelinepenalty=10000\footnote{One might worry that binning the data yields additional sources of bias not accounted for in \eref{eq:TemplateBiasImp}. As long as $h_b \ll h$ and linear interpolation is applied, however, this additional bias is negligible. In practice, we find that $h_b$ as large as $h/2$ still yields a consistent estimate. Of course, it is preferable to take $h_b$ as small as allowed by memory limitations.}} 

We use the bootstrap \cite{efron79, scott, hall1995bootstrap, horowitz2001bootstrap} to estimate the variance. The flexibility of the bootstrap is particularly well-suited to our case, where the variance needs to be propagated through complicated kinematic cuts. For a given statistic $\epsilon$ (\emph{e.g.}, $\epsilon$ can be the efficiency for the cuts of interest) and estimator $\hat{\epsilon}$, the bootstrap relies on creating a set of resamples of the observed dataset. Each resampled dataset then yields a value for  $\hat{\epsilon}$, the distribution of which is used to measure properties of the estimator $\hat{\epsilon}$ (\emph{e.g.},~its variance). 

To compute the variance, one first converts a given dataset $\big\{\vec{z}_i \big \}$ into a set of bin counts $\{n_{\vec q}\}$. One then fluctuates each element of the set $\big \{ n_{\vec q} \big \}$ a total of $\nbs$ times, thus creating $\nbs$ fluctuated datasets, denoted by $\big \{ n_{\vec q} \big \}_b$, with $b\in \{1,..., \nbs\}$.  Each element of $\big \{ n_{\vec q} \big \}_b$ is sampled from the Poisson distribution:\footnote{For $N_T$ data points $\big \{\vec{z}_i \big \}$ the standard bootstrap procedure generates each resampled dataset by drawing exactly $N_T$ samples from the empirical distribution (\emph{i.e.},~from the observed dataset) with replacement.  Our Poisson fluctuation procedure (which we choose for computational convenience) is not identical to the standard bootstrap, but in practice we find that it provides robust estimates of the variance.  The accuracy of this prescription deserves further scrutiny in any future experimental study.} 
\be
\big\{n_{\vec q}\big\}_{b}\sim \textrm{Poisson}\big(n_{\vec q}\big) \, .
\label{eq:A_n_fluctuate}
\ee
For each fluctuated dataset, one can compute a corresponding $\left\{\hat{\rho}\big(\vz_{\,}\big)\right\}_{b}$ using the same binned kernel smoothing procedure as in \eref{eq:binnedRhoHat}: 
\be
\left \{ \hat{\rho}\big(\vz_{\,}\big) \right \}_b =\frac{\kappa_b}{\mathcal{N}_b} \sum_{\vec q} \big\{n_{\vec q} \big\}_b \, K_h\big(\vec q\,'-\vec q\big),
\label{eq:A_rho_fluctuate}
\ee
where, as before, $\kappa_{b}$ is a normalization constant, and $\mathcal{N}_b$ is the total number of events in the $b^{\rm{th}}$ fluctuated dataset.  Finally, we obtain an estimate of the smoothing variance:
\be
\hat{v}^2\big(\vz_{\,}\big)=
\frac{1}{\nbs-1}\sum_{b=1}^{\nbs} \Big[ \big\{\hat\rho\big(\vz_{\,}\big)\big\}_b - \Big\langle \big\{ \hat\rho\big(\vz_{\,}\big) \big \} \Big\rangle_{\text{\tiny{BS}}} \Big]^2\,; \qquad 
\Big\langle \big\{ \hat\rho\big(\vz_{\,}\big) \big \} \Big\rangle_{\text{\tiny{BS}}} = \frac{1}{\nbs}\sum_{b=1}^{\nbs} \big\{\hat\rho\big(\vz_{\,}\big)\big\}_b\,.
\label{eq:TemplateVarImp}
\ee
In practice, $\nbs\simeq 100$ is sufficient for a reasonable estimate of the variance.  

For our application, it is important that density estimates $\hat{\rho}(\vec z)$ be associated with a confidence interval.  However, due to the presence of asymptotic bias (see especially the discussions in \cite{hall1995bootstrap, horowitz2001bootstrap}), a naive application of the smoothing variance in \eref{eq:TemplateVarImp} might lead us to construct incorrect confidence intervals that are centered at $\langle \hat{\rho}(\vec z)\rangle$ instead of at ${\rho}(\vec z)$. There are two approaches to this problem, both of which can be used to ensure the construction of properly centered confidence intervals. The first is to deliberately undersmooth the data so that bias is negligible in comparison to the root variance. The second is to explicitly correct for the bias. In practice, we find that it is necessary to use both approaches simultaneously. Relying solely on undersmoothing is inadequate because background estimates rely on estimates of $\rho(\vz)$ at a variety of different values of $\vz$; to ensure small bias over the entire range of relevant values of $\vz$ would lead to too much undersmoothing. Relying solely on bias correction is inadequate because of the difficulty of constructing an accurate bias estimator. By making use of both approaches simultaneously, it is possible to mitigate the problems of asymptotic bias.  In practice, for any specified cut, we choose a bandwidth such that $ \hat\sigma_V \gtrsim 2\, \hat\sigma_B$, where $\hat\sigma_B$ is a measure of the total bias of $\hat N_e^\star[C]$, see \eref{eq:EventBias}.  It is worth stressing the need to pay careful attention to these issues, because they play a central role in determining the robustness of the final statistical uncertainties associated with the template procedure.  Note that more sophisticated approaches that, \emph{e.g.} make use of adaptive kernel smoothing or implement bias correction
at the level of $\hat\epsilon[C]$, may allow for an improved treatment of asymptotic bias. 

To introduce explicit bias correction, we define the bias-corrected density estimate $\hat{\rho}^\star(\vec z)$:
\be
\hat{\rho}^\star\big(\vz_{\,}\big) \equiv \hat{\rho}\big(\vz_{\,}\big) - \hat{b}\big(\vz_{\,}\big) \, .
\label{eq:A_rho_star}
\ee
Note that $\int  \dd^D z \,\srho \big(\vz_{\,}\big)=1$ as long as $\hat{b}(\vec z)$ is the difference of two probability distributions as computed in \eref{eq:TemplateBiasImp}.  The resulting $\hat{\rho}^{\star}$ is not necessarily positive everywhere.  We find that this does not cause any problems for the cut efficiencies computed in this paper, but one should be aware of this issue.  \eref{eq:TemplateVarImp} can also be used to define the bias-corrected variance via the substitution\footnote{The variance itself is not bias-corrected, but rather $v^{2\star}$ represents the dispersion about the bias-corrected density estimate $\srho(\vec z)$. }
\be
\trho \rightarrow \srho \;\;\Longrightarrow \;\;v^2 \rightarrow v^{2\star}.
\ee

Using the bias-corrected $\hat{\rho}^\star(\vec z)$, one can construct a bias-corrected estimator $\hat{\epsilon}^{\star}$, which has substantially reduced bias. Note that $\epsilon^{\star}$ is still a biased estimator, but the bias is now of higher order in the bandwidth $h$. The cost of reducing bias is an increase in the variance, which can be reliably computed with the bootstrap.  

Given the ensemble of bias-corrected estimators, the corresponding statistical error is computed with the bootstrap.  As discussed in Sec.~\ref{sec:PredictionsWithErrorBars}, each of the $\left\{\srho\right\}_b$ is used as an input to generate an ensemble of bootstrapped results $\big\{N^{\star}_e[C]_1, ... , N^{\star}_e[C]_{\nbs} \big \}$ for a given set of cuts $C$.  The variance of the final prediction is estimated using this ensemble:
\be
\hat\sigma_V^2 \simeq
\frac{1}{\nbs-1}\sum_{b=1}^{\nbs}\Big(\mathcal{N}^{\star}_{e}[C]_b- \left < \mathcal{N}^{\star}_{e}[C] \right >_\text{\tiny{BS}}\Big)^2 \quad \text{where} \quad
\left < \mathcal{N}^{\star}_{e}[C] \right >_\text{\tiny{BS}} =\frac{1}{\nbs}\sum_{b=1}^{\nbs} \mathcal{N}^{\star}_{e}[C]_b \, .
\label{eq:sigmaVApp}
\ee
This is to be distinguished from the variance of the templates themselves, $v^2(\vz)$.

The next subsection presents a validation study showing that the bias correction and variance estimates outlined above are statistically robust and lead to sound confidence intervals.

\subsection{Validation}
\label{sec:A_validation}
\begin{figure}[h!] 
   \centering
   \includegraphics[width=0.48\textwidth]{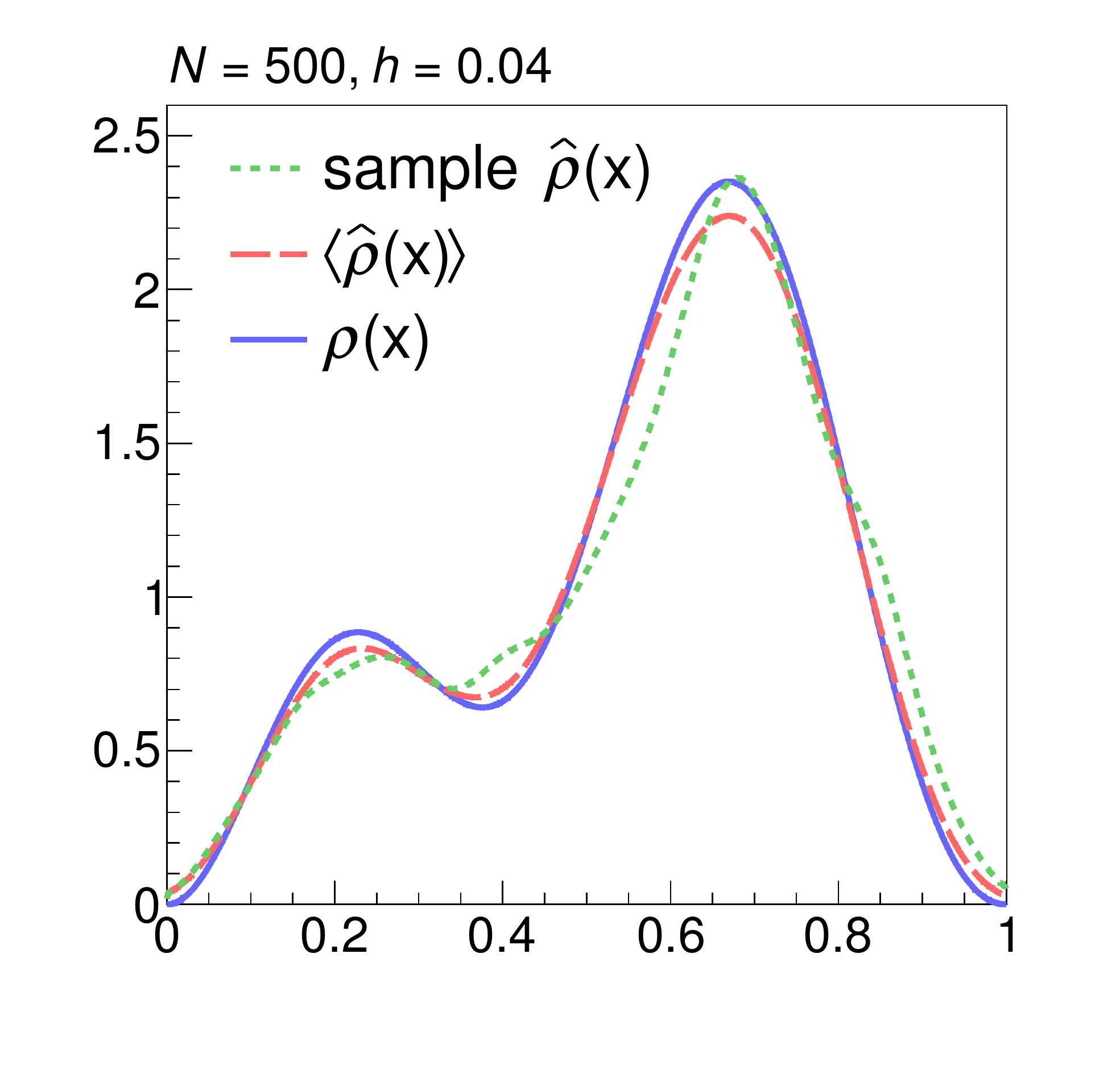} 
   \includegraphics[width=0.48\textwidth]{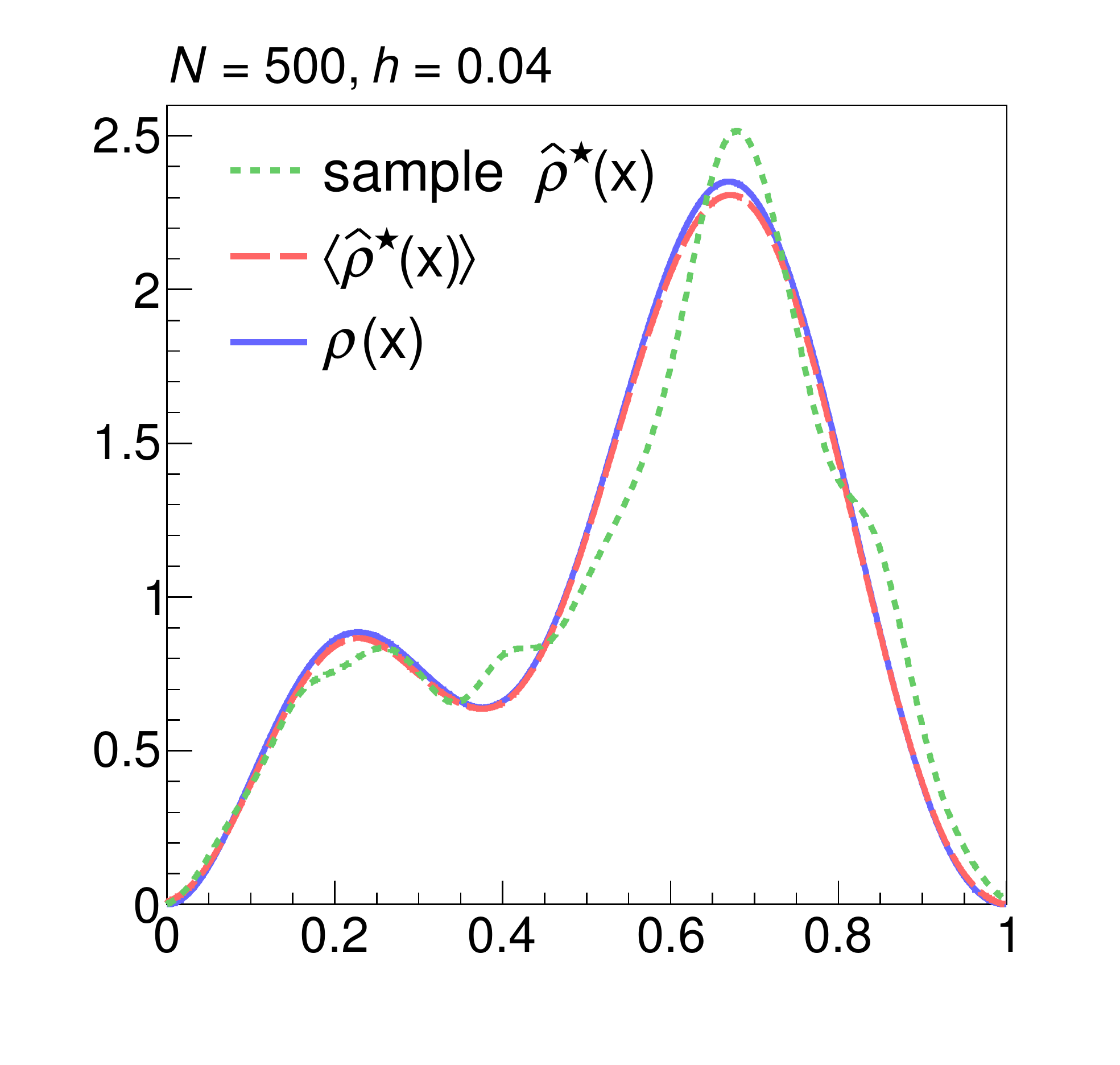} 
   \caption{Results of the first validation test with 1000 trial experiments.  Left: $\rho(x)$ is the true distribution (blue), $\hat\rho(x)$ is the kernel estimate from a single trial experiment (green), and $\langle\hat\rho(x)\rangle$ is the average over all experiments (red). Right: The bias-corrected results for the same set of trial experiments.}
   \label{fig:A_rho}
\end{figure}

This subsection presents a validation study for our particular implementation of kernel smoothing.  The first test evaluates the accuracy of the estimates $\trho(\vz)$ by applying kernel smoothing to a known one-dimensional probability distribution $\rho(x)$ with $x\in [0,1]$.  The second test evaluates the data-driven procedure for estimating cut efficiencies by applying cuts to the joint probability distribution $\rho(x_1,x_2)=\rho(x_1)\rho(x_2)$. Both tests use the distribution $\rho(x)\sim x(1-x)(\cos(8x)+1.5)$, illustrated in Fig. \ref{fig:A_rho}.

The first test is formulated as follows:
\begin{enumerate}
\vspace{-5pt}
\item Do $N_\text{trials} = 1000$ trial experiments, with each experiment containing $N=500$ events, $\{x_1, x_2, ..., x_N \}$, drawn from $\rho(x)$. 
\vspace{-8pt}
\item For each experiment $i$, compute the estimates $\hat\rho_i(x),\hat\rho_i^{\star}(x),\hat{b}_i(x)$ and $\sqrt{\hat{v}^{2\star}_i(x)}$ using the methods outlined in App.~\ref{sec:A_implementation}.
\vspace{-8pt}
\item Compute the averages $\langle\hat\rho(x)\rangle,\langle\hat\rho^\star(x)\rangle,\langle\hat{b}(x)\rangle$ and $\langle\sqrt{\hat{v}^{2\star}(x)}\rangle$ from the 1000 experiments.
\vspace{-22pt}
\item Using the true distribution $\rho(x)$, compute
${b}(x)$ and $\sqrt{v^{2\star}(x)}$. 
\end{enumerate}
The resulting kernel estimates are illustrated in Fig. \ref{fig:A_rho}. The bandwidth is chosen with Silverman's Rule-of-Thumb with a moderate amount of undersmoothing ($c\simeq 0.9 \, c_{\text{\tiny{AMISE}}}$ and $h\simeq 0.04$ in \eref{eq:A_silverman}). The bin width is $h_b=0.0025 \ll h$. The bias-corrected version on the right shows substantial reduction in the average bias 
(\emph{i.e.},~red matches blue). The variance, however, increases as can be seen by the increased error of the sample estimate, shown in green. 

\begin{figure}[h!] 
   \centering
   \includegraphics[width=0.48\textwidth]{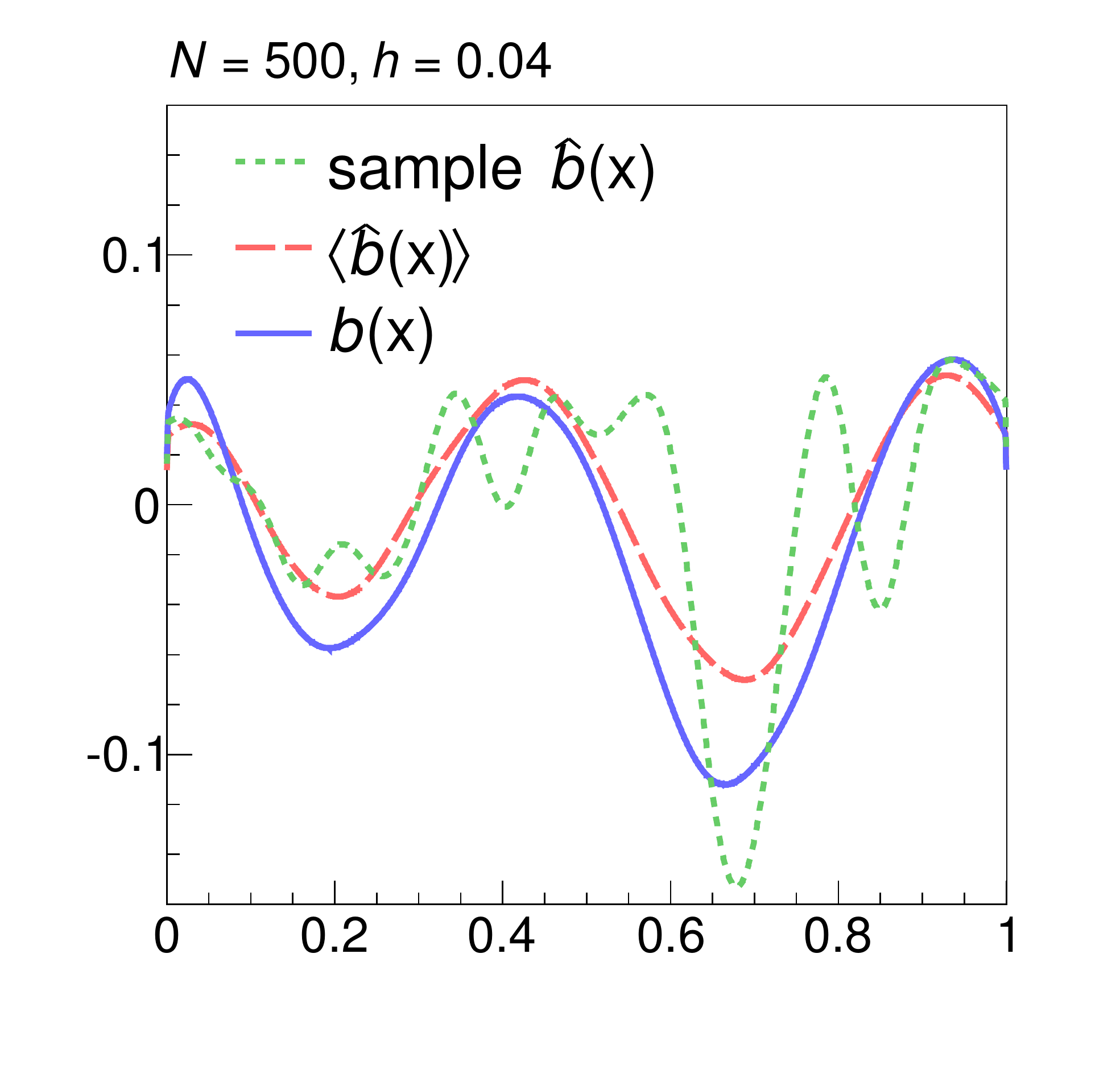} 
   \includegraphics[width=0.48\textwidth]{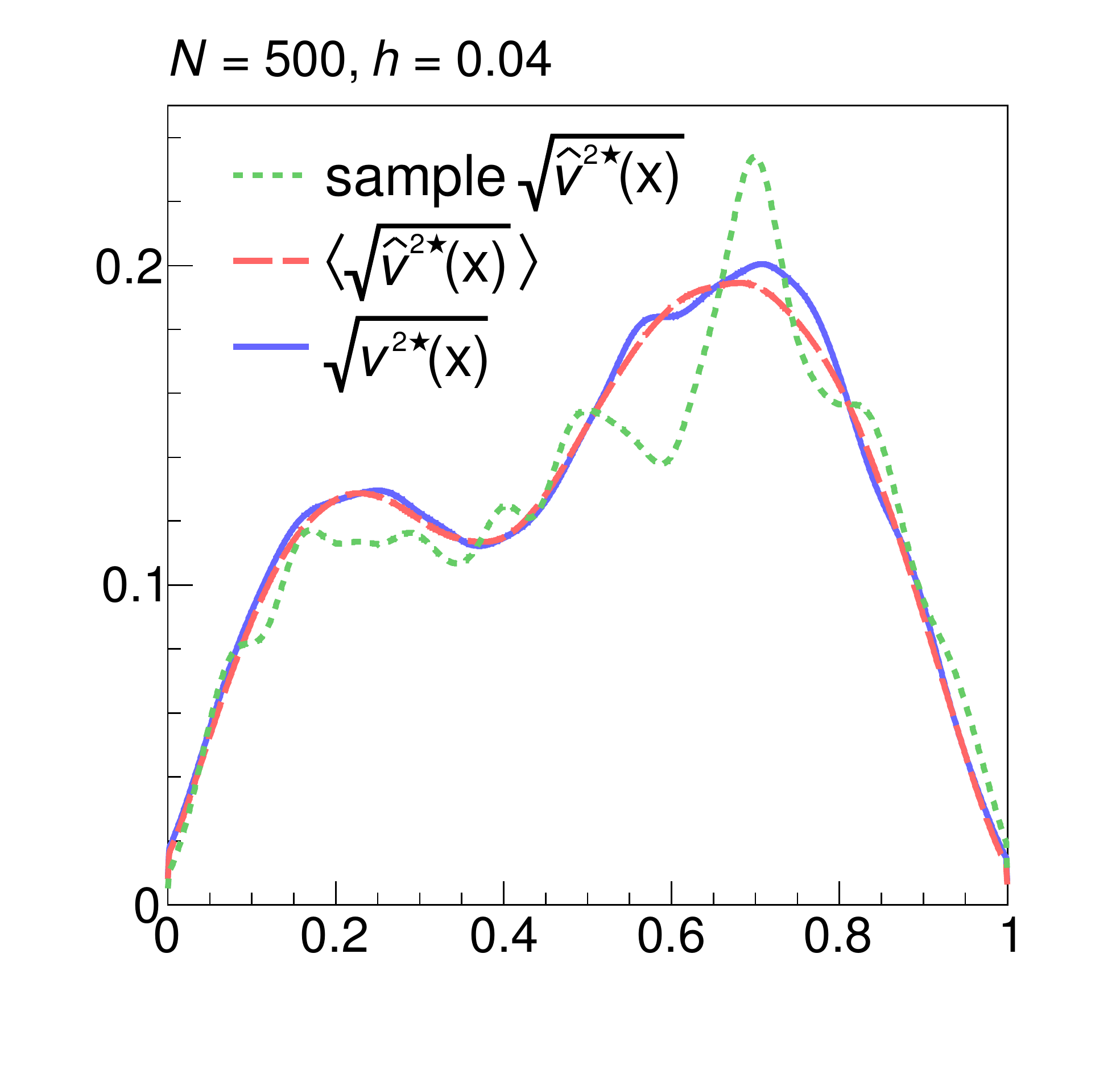} 
   \caption{The bias and variance for the first validation test. Left:  $b(x)$ is the true bias (blue), $\hat{b}(x)$ is the bias estimate for a single trial experiment (green), and $\langle \hat{b}(x) \rangle$ is the estimated bias averaged over all trial experiments (red). Right: The corresponding results for the square root of the bias-corrected variance, $\sqrt{v^{2\star}(x)}$. }
   \label{fig:A_bias_var}
\end{figure}

The performance of the bias and variance estimators is shown in Fig. \ref{fig:A_bias_var}.  The left panel shows systematic underestimation of the bias. The bias, however, is subdominant to the variance as a consequence of undersmoothing.  Since the bias is generally of the correct sign and order-of-magnitude, bias correction leads to confidence intervals that have improved probability coverage (as we will see explicitly below).  Compared to the bias, the variance can be more reliably estimated, as seen in the right panel.  It is likely that the bias estimator can be improved, which would be an important goal for any experimental study. 

The second test goes a step further and allows us to evaluate the statistical performance of our procedure for estimating cut efficiencies. In particular it constitutes a test of the smoothing variance $\hat\sigma_V^2$ and the probability coverage of the corresponding confidence interval.  It differs from the cut computations in Sec.~\ref{sec:Applications} only in that: i) the kinematic sample is effectively of size 1 and ii) the dimension is $D=1$.  

The test is formulated as follows:
\begin{enumerate}
\item Do $N_\text{trials}=5000$ trial experiments, with each experiment containing $N=500$ events, $\{x_1, x_2, ..., x_N \}$, drawn from $\rho(x)$.
\vspace{-8pt}
\item For each trial experiment $i$, compute $\hat{\rho}_i(x)$ and $\hat{\rho}_i^\star(x)$. \\Let \mbox{$\epsilon [C]_i\equiv \int_{C} \mathrm{d}x_1 \mathrm{d}x_2\, \hat{\rho}_i(x_1)\hat{\rho}_i(x_2)$}, where $C$ is given by the region $x_1+x_2> x_{\textrm{cut}}$. Compute $\hat\epsilon[C]_i$
and  $\hat\epsilon^\star[C]_i$.
\vspace{-8pt}
\item For each trial experiment $i$, calculate the bias estimator $\hat\sigma_{Bi}=\hat{\epsilon}[C]_i-\hat{\epsilon}^\star[C]_i$, as
in \eref{eq:EventBias}, and use the bootstrap to calculate $\hat\sigma_{Vi}$, as in \eref{eq:EventVar}. 
\vspace{-8pt}
\item Using the true distribution $\rho(x)$, compute $\epsilon[C]$ and form the statistics
\be
s_i\equiv \frac{\hat\epsilon[C]_i-\epsilon[C]}{\hat\sigma_{Vi}} \qquad \text{and} \qquad
s_i^\star\equiv \frac{\hat\epsilon^\star[C]_i-\epsilon[C]}{\hat\sigma_{Vi}}\, .
\label{eq:teststatistics}
\ee
Note that $\hat\sigma_{Vi}$ does not include a bias correction when constructing $s_i$.
\vspace{-8pt}
\item Finally, calculate the variance of the estimates $\{\hat\epsilon^\star[C]_i\}$:
\be
\sigma_V^2 = \frac{1}{N_\text{trials}-1}\sum_{i=1}^{N_\text{trials}} (\hat\epsilon^\star[C]_i-\langle\epsilon^\star[C]\rangle_\text{trials})^2\qquad
\langle\epsilon^\star[C]\rangle_\text{trials}= \frac{1}{N_\text{trials}}\sum_{i=1}^{N_\text{trials}} \hat\epsilon^\star[C]_i \, .
\ee
\end{enumerate}

\begin{figure}[h!] 
   \centering
   \includegraphics[width=0.48\textwidth]{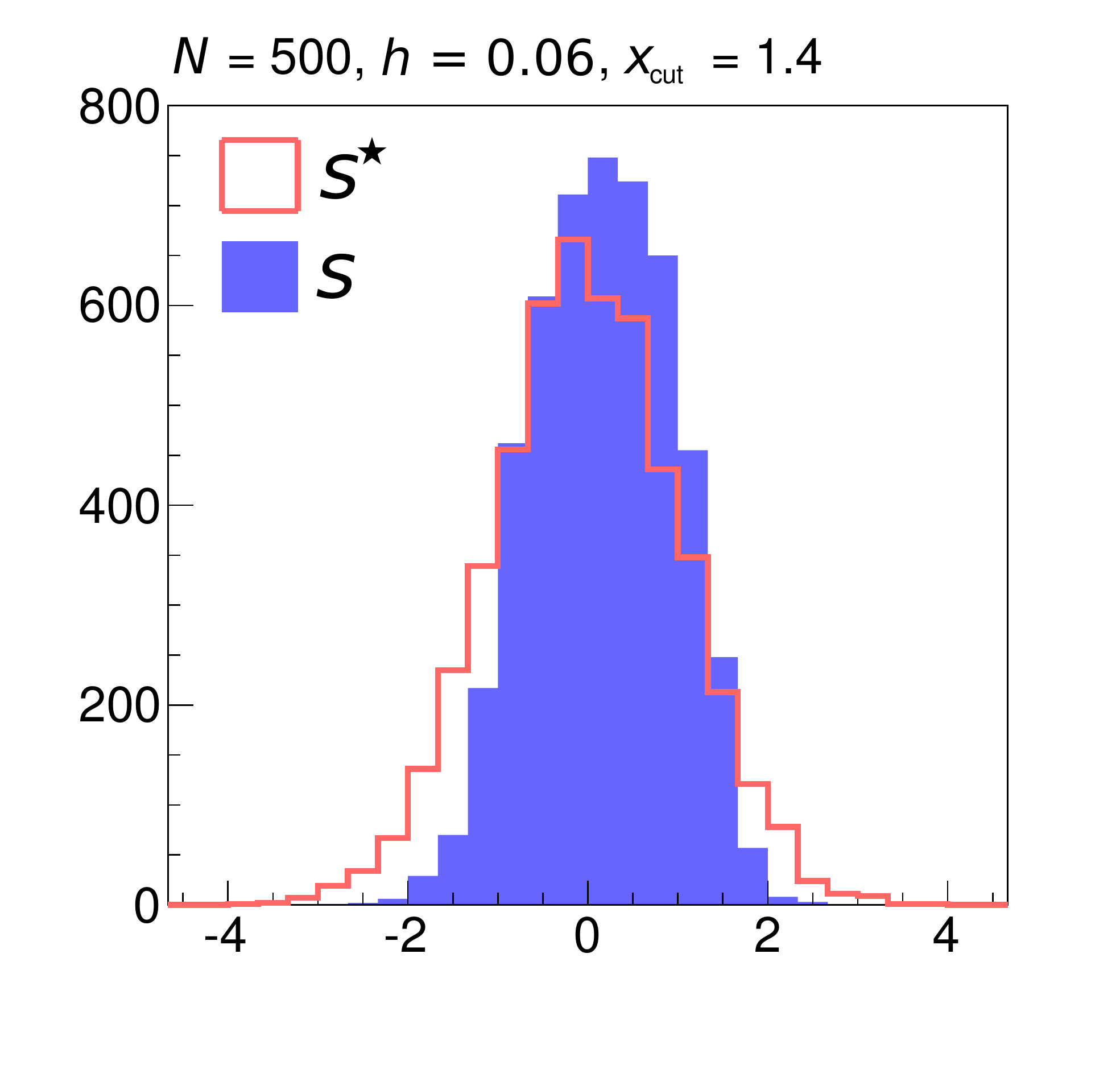} 
   \includegraphics[width=0.48\textwidth]{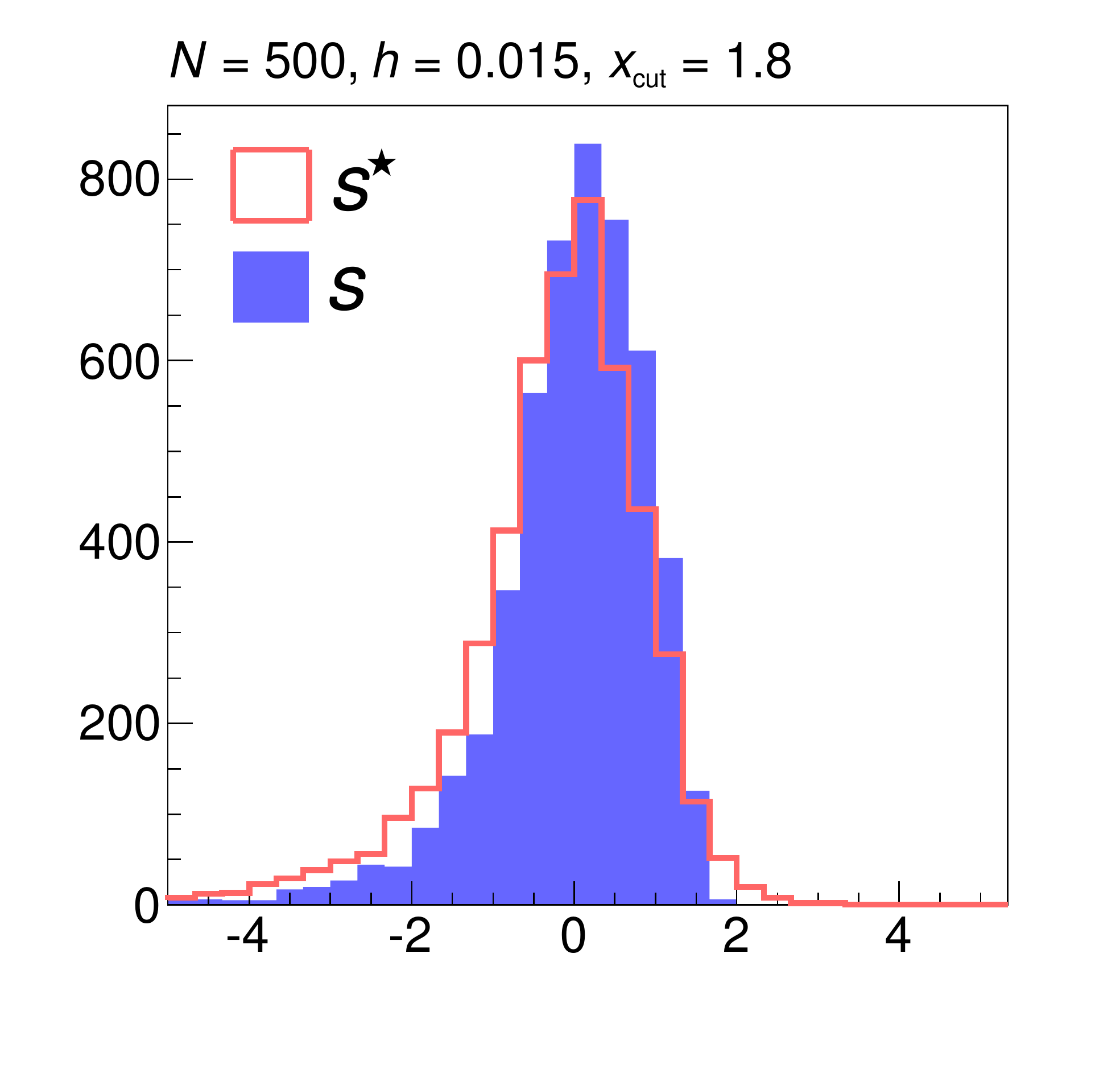} 
   \caption{The results of the second validation test, which evaluates the probability coverage of the confidence intervals corresponding to the cut efficiencies $\hat\epsilon$ and $\hat\epsilon^\star$ with $x_{\textrm{cut}}=1.4$ (left) and $1.8$ (right).  The statistics $s$ and $s^\star$ are defined in \eref{eq:teststatistics}.  The fact that $s^\star$ is approximately Gaussian with mean zero and unit variance indicates that the corresponding confidence intervals are sound. }
   \label{fig:A_relative_sig}
\end{figure}

The distributions for $s$ and $s^\star$ are shown in Fig.~\ref{fig:A_relative_sig}.  Note that a robust confidence interval should result in a statistic $s$ ($s^\star$) that is approximately Gaussian with zero mean and unit variance. As can be seen in the figure, bias correction significantly improves the Gaussianity of $s^\star$ (which corresponds to the bias-corrected estimate $\hat\epsilon^\star$)   as compared to $s$ (which corresponds to the estimate $\hat\epsilon$). For $x_{\textrm{cut}}=1.4$ and with $h=0.06$ the bias is roughly a factor of 20 smaller than the variance, but the difference between $s$ and $s^\star$ is still non-negligible.  For $x_{\textrm{cut}}=1.8$, $h$ needs to be significantly smaller with $h=0.015$ in order to achieve a subdominant bias. The bias in this case is roughly a factor of two smaller than the variance, and the bias correction results in a sizable shift from $s$ to $s^\star$.  For both cuts, bias correction improves the probability coverage of the confidence interval for $\hat\epsilon^\star$ (especially its centeredness) and thus justifies the error bars given for the background estimates in Sec.~\ref{sec:Applications}.  

\begin{figure}[h!] 
   \centering
   \includegraphics[width=0.48\textwidth]{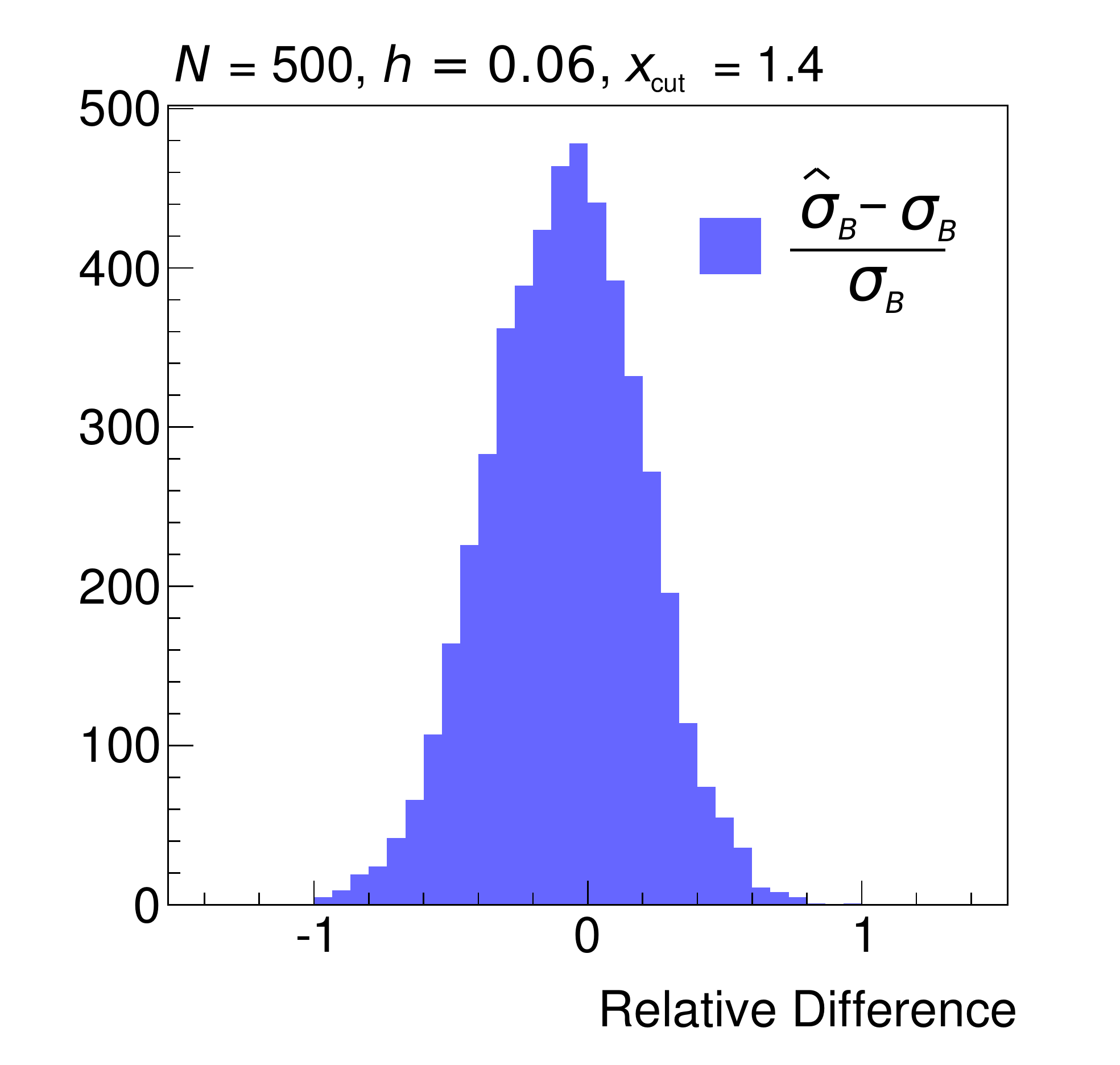} 
   \includegraphics[width=0.48\textwidth]{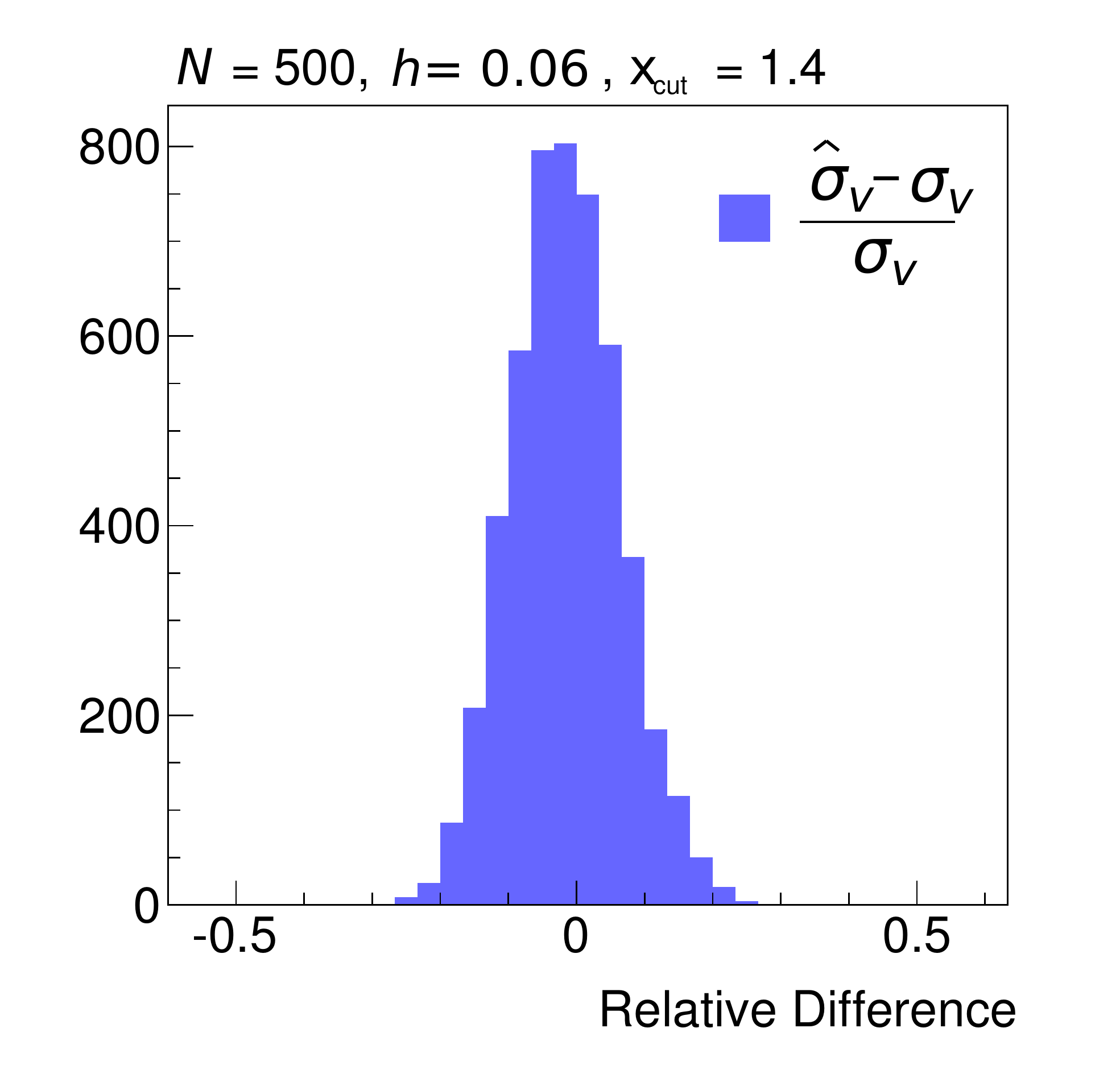} 
   \caption{The performance of $\hat\sigma_B$ and $\hat\sigma_V$ in the 5000 trial experiments of the second validation test.  Plotted is the distribution of deviations for the estimator $\hat\sigma_B=|\hat{\epsilon}[C]-\hat{\epsilon}^\star[C]|$ from the true bias $\sigma_B = |{\hat\epsilon}[C]-{\epsilon}[C]|$ (left).  Also shown is the deviation of the estimator $\hat\sigma_V$ from the true root variance $\sigma_V$ (right).}
\label{fig:A_relative_eff_var}
\end{figure}

Finally, the performance of the estimators $\hat\sigma_B$ and $\hat\sigma_V$ is shown in Fig.~\ref{fig:A_relative_eff_var}.  As is apparent, variance estimation is much more reliable than bias estimation.  The bias estimator $\hat\sigma_B$ (as opposed to the bias estimator $\hat b(\vz)$ used to define $\srho(\vz)$) need not be very accurate because it is only used to ensure that we are in a statistical regime where the bias is subdominant so that confidence intervals have good probability coverage.  Note that the performance of $\hat\sigma_B$ illustrated in the figure is for the particular cut value of $x_{\textrm{cut}}=1.4$; the agreement between $\hat\sigma_B$ and $\sigma_B$ varies as a function of $x_{\textrm{cut}}$, although the overall level of agreement shown here is representative.

\section{QCD Monte Carlo}
\label{sec:MCFramework}

This section describes the simulated data utilized in our studies.  Up to four final-state partons were simulated at the generator-level, which were then merged with a shower.  The final hadron-level output was pixelated to model the finite granularity of the calorimeter.\footnote{This more closely resembles the setup for ATLAS than for CMS, where particle-flow jets are used instead.  We expect that this will have little to no impact on our results.}  Finally, these pixels were clustered into fat anti-k$_T$ jets with $R=1.0$.  This level of detail in the simulation is sufficient for demonstrating the viability of our data-driven technique.  Note that if we were
considering backgrounds for searches involving missing energy, it would be important to also simulate detector-smearing effects.

In more detail, parton-level QCD events were generated using \texttt{Madgraph5} version 5.1.5.10 at $\sqrt{s}=8$ TeV and merged with the parton shower via the MLM ``MadGraph style" prescription using \texttt{PYTHIA} version 8.1.76 \cite{Sjostrand:2007gs}.  To more accurately model the large phase space for multijet production at the LHC, we employed a variation of the weighting technique discussed in \cite{Avetisyan:2013onh}.  Specifically, events were binned in both jet multiplicity (2 exclusive, 3 exclusive, and 4 inclusive jets) and in $H_{T}$, see Table \ref{table:statistics}. 

\begin{table}[t]
\renewcommand{\arraystretch}{1.5}
\setlength{\tabcolsep}{5.2pt}
\begin{center}
\begin{tabular}{|c|c||c|c||c|c||c|c|}
\hline
\multirow{2}{*}{bin} & \multirow{2}{*}{$H_{T}$ [GeV]}  &  \multicolumn{2}{c||}{$2\,j$} &   \multicolumn{2}{c||}{$3\,j$}  &   \multicolumn{2}{c|}{$4\,j$} \\
\cline{3-8}
& & $\sigma$ [pb] & $N_\mathrm{events}$ &  $\sigma$ [pb] & $N_\mathrm{events}$ &  $\sigma$ [pb] & $N_\mathrm{events}$ \\
\hline
1  & $0 - 50$ & $\scinot{9.0}{7}$ & $\scinot{2.8}{6}$  & 0 & 0  & 0 & 0 \\
2  & $50 - 100$ & $\scinot{1.0}{8}$ & $\scinot{4.1}{6}$  & $\scinot{6.8}{6}$ & $\scinot{3.0}{6}$  & $\scinot{3.2}{4}$ & $\scinot{8.3}{5}$ \\
3  & $100 - 200$ & $\scinot{4.3}{6}$ & $\scinot{3.0}{6}$ & $\scinot{6.5}{6}$ & $\scinot{3.1}{6}$  & $\scinot{2.8}{6}$ & $3.3\times10^6$ \\
4  & $200-400$ & $\scinot{1.1}{5}$ & $\scinot{1.6}{6}$ & $\scinot{2.9}{5}$  & $\scinot{1.8}{6}$  & $\scinot{1.2}{6}$ & $4.2\times10^6$ \\
5  & $400-800$ & $\scinot{1.9}{3}$ & $\scinot{8.2}{5}$ & $\scinot{6.9}{3}$ & $\scinot{8.4}{5}$  & $\scinot{9.5}{4}$ & $5.2\times10^6$ \\
6  & $800-1400$ & $\scinot{2.7}{1}$ & $\scinot{4.9}{5}$  & $\scinot{1.1}{2}$ & $\scinot{4.8}{5}$  & $\scinot{3.6}{3}$ & $4.9\times10^6$ \\
7  & $1400-2000$ & $\scinot{6.3}{-1}$ & $\scinot{4.0}{5}$  & $\scinot{2.7}{0}$ & $\scinot{3.8}{5}$ & $\scinot{1.3}{2}$ & $5.1\times10^6$ \\
8  & $2000-3000$ & $\scinot{4.5}{-2}$ & $\scinot{3.9}{5}$  & $\scinot{1.9}{-1}$. & $\scinot{3.8}{5}$  & $\scinot{1.0}{1}$ & $5.4\times10^6$ \\
9  & $3000-4000$ & $\scinot{1.0}{-3}$ & $\scinot{4.2}{5}$  & $\scinot{4.4}{-3}$ & $\scinot{4.2}{5}$  & $\scinot{2.2}{-1}$ & $4.6\times10^6$ \\
10  & $4000-5000$ & $\scinot{2.6}{-5}$ & $\scinot{4.3}{5}$  & $\scinot{1.1}{-4}$ & $\scinot{4.6}{5}$  & $\scinot{5.2}{-3}$ & $5.4\times10^6$ \\
11  & $5000-6000$ & $\scinot{4.4}{-7}$ & $\scinot{4.9}{5}$  & $\scinot{1.9}{-6}$ & $\scinot{5.6}{5}$  & $\scinot{7.8}{-5}$ & $4.8\times10^6$ \\
12  & $ > 6000$ & $2.7\times10^{-9}$ & $\scinot{5.7}{5}$  & $\scinot{1.2}{-8}$ & $\scinot{6.9}{5}$  & $\scinot{3.8}{-7}$ & $4.2\times10^6$ \\
\hline
\end{tabular}
\caption{The number of MC events after matching in the different $H_{T}$ and jet multiplicity bins.  The associated matched cross sections for each bin are also provided.}
\label{table:statistics}
\end{center}
\end{table}

The post-\texttt{PYTHIA} hadron-level information was then passed to a code based on \texttt{FastJet} version 3.0.2  \cite{Cacciari:2011ma,Cacciari:2005hq}.  The granularity of the calorimeter was simulated by creating a grid of cells in the $\eta$-$\phi$ plane with size $0.1\times 0.1$ for $|\eta| < 2.5$ and $0.2\times 0.2$ for $2.5 < |\eta| < 4.5$; any particles with $|\eta| > 4.5$ were discarded.  For each event, the energy of all particles that fell within that cell was summed.  This gave the energy of the cell for the given event.  The energy and position of these cells were combined into light-like four-vectors that were used as inputs to the \texttt{FastJet} routines.  

Next, the cells were clustered into $R=1.0$ anti-$k_T$ jets.  In order to reduce sensitivity to the impact of pileup, the jets were groomed using the trimming algorithm with $f_{\textrm{cut}}=0.05$ and $R_{\textrm{trim}}=0.3$.  The fat jets used in our study are the output of this trimming procedure.  $N$-subjettiness was computed using the \texttt{FastJet} package \cite{Thaler:2010tr, Thaler:2011gf} with the `\texttt{min\_axes}' algorithm and $\beta = 1$. 

One of the main goals of Sec.~\ref{sec:Applications} is to demonstrate that the error bars associated with the smoothing procedure are reasonably small.  As discussed above, the width of the kernel scales with the number of events in the training sample, $N_T$, and the total AMISE scales as ${N_T}^{-4/(D+4)}$ (see \eref{eq:AMISE_Final}).  As a result, we must understand the size of the training
sample available at the LHC in order to make realistic projections for the statistical uncertainties of the template procedure.  For very inclusive triggers, the event rate can be so large that it is not possible to write all events that pass selection cuts to tape.  To deal with this, the collaborations perform prescaling, \emph{i.e.}~they record only a fraction of the events and reweight them to account for events that went unrecorded.  Prescaling is relevant because the weighted events come with larger statistical uncertainties than the naive expectation of $\sqrt{N}$.

The training samples we consider are taken from low-multiplicity exclusive events.  We assume these events are obtained through a dijet trigger, which requires an enormous prescaling factor for low-$p_{T}$ cuts. To avoid these issues, we choose a $p_{T}$ cut high enough that prescaling is not necessary. The $p_{T}$ threshold is computed as follows:
\begin{enumerate}
\item For a given threshold $p_{T}$, compute the two fat jet cross section;
\item Assume that the instantaneous luminosity at the LHC is $10^{34}$ cm$^{-2}$/s, and compute the prescaling factor required to keep the trigger rate below 200 Hz;
\item Choose the lowest $p_{T}$ threshold such that the prescaling factor is roughly 1. 
\end{enumerate}
This procedure yields a $p_{T}$ threshold of 200 GeV.  Since both our mock analyses stay above this $p_T$ threshold, 
it is not necessary for us to take prescaling explicitly into account. 
\end{spacing}
 
\pagebreak

\begin{spacing}{1.1}
\bibliography{DataDriven}
\bibliographystyle{utphys}
\end{spacing}
\end{document}